# Mechanics of Stabbing: Biaxial Measurement of Knife Stab Penetration of Skin Simulant

M. D. Gilchrist, S. Keenan, M. Curtis, M. Cassidy, G. Byrne & M. Destrade

2007

## ABSTRACT

This paper describes the development and use of a biaxial measurement device to analyse the mechanics of knife stabbings. In medicolegal situations it is typical to describe the consequences of a stabbing incident in relative terms that are qualitative and descriptive without being numerically quantitative. Here, the mechanical variables involved in the possible range of knife-tissue penetration events are considered so as to determine the necessary parameters that would need to be controlled in a measurement device. These include knife geometry, in-plane mechanical stress state of skin, angle and speed of knife penetration, and underlying fascia such as muscle or cartilage. Four commonly available household knives with different geometries were used: the blade tips in all cases were single-edged, double-sided and without serrations. Appropriate synthetic materials were used to simulate the response of skin, fat and cartilage, namely polyurethane, compliant foam and ballistic soap, respectively. The force and energy applied by the blade of the knife and the out of plane displacement of the skin were all used successfully to identify the occurrence of skin penetration. The skin tension is shown to have a direct effect on both the force and energy for knife penetration and the depth of out of plane displacement of the skin simulant prior to penetration: larger levels of in-plane tension in the skin are associated with lower penetration forces, energies and displacements. Less force and energy are also required to puncture the skin when the plane of the blade is parallel to a direction of greater skin tension than when perpendicular. This is consistent with the observed behaviour when cutting biological skin: less force is required to cut parallel to the Langer lines than perpendicularly and less force is required to cut when the skin is under a greater level of tension. Finally, and perhaps somewhat surprisingly, evidence is shown to suggest that the quality control processes used to manufacture knives fail to produce consistently uniform blade points in knives that are nominally identical. The consequences of this are that the penetration forces associated with nominally identical knives can vary by as much as 100%.

# INTRODUCTION

Stabbing is the most common cause of homicide in the UK and Ireland [1] and a question often asked of a medical witness is: what was the degree of force involved? The answer is considered significant in determining an alleged assailant's intent to cause harm. The magnitude of force used in a stabbing incident can be difficult to quantify except in relative terms and pathologists describe a stabbing force in relative terms that range from mild through moderate and considerable to severe [1]. A mild level of force would typically be associated with penetration of skin and soft tissue whereas moderate force would be required to penetrate cartilage or rib bone. Severe force, on the other hand, would be typical of a knife impacting dense bone such as spine and sustaining visible damage to the blade. A quantitative measure for the force required to penetrate different tissues and protective clothing would avoid the need for a subjective scale and may provide insight into the much used defence that the victim 'ran onto the knife!' This paper describes the development of a device to accurately and repeatedly measure the penetration characteristics of knives stabbing soft deformable skin-like materials.

Skin can be divided structurally into two layers: the outer epidermis and the underlying dermis, which provides most of the strength. The physical properties of skin vary with age, health, body weight, location on a body and orientation. Skin thickness can range from 4mm to 0.5mm and it has widely different mechanical properties on different parts of the body. The epidermis is a thin exterior covering composed of cells and cellular debris. Its main function is to protect the underlying skin layers and it has limited mechanical strength. The dermis is a matrix of alignable collagen fibres (c. 35% by volume) and elastin fibres (c. 0.4% by volume) which are interwoven in a ground substance of proteoglycans, water and cells [2]. It is the dermal layer that provides most mechanical strength and is capable of withstanding large deformations. The strength is due to collagen fibres which are almost inextensible and fail at strains of 5-6% [3] and strengths in the range of 147-343 MPa, varying with position in the body [4]. The elastin, on the other hand, is highly deformable, and provides the dermis with its high elasticity. As a person ages, the elastin in the skin degrades and the cross linking of collagen fibres increases: this increase in cross linking reduces the amount of water stored in the skin and this is the primary cause of wrinkling. Older skin with a reduced elastin content and increased collagen cross linking will have reduced elasticity and its failure strain will also be reduced.

Tensile tests on skin produce a nonlinear stress-strain relationship that is 'J'-shaped [5]. When the dermis is in its normal relaxed state and not subjected to strain, the collagen and elastin fibres are not highly ordered. As the skin is strained, the elastin fibres start to carry the load while the collagen fibres are still disorganised and unstrained. At increased levels of strain, the collagen fibres are increasingly aligned in the direction of the load and start to share a larger proportion of the load. At the highest levels of strain possible, the collagen fibres carry almost all the load because collagen is less extensible than elastin. An increase in force in this region of the stress-strain curve gives a much smaller extension than in the previous regions.

Throughout most of a body, the skin is in a state of unequal biaxial tension. The state of tension in the skin is due to the movement of joints local to the skin and to the volume of mass under it [6]. On a body at rest, the local directions of maximum tension are collinear with the local orientations of collagen fibres within the skin and these directions are known as Langer lines or cleavage lines. The influence of this unequal state of biaxial tension is manifest if incisions are made parallel or perpendicular to Langer lines: a parallel incision will gape open less than one perpendicular and will heal with a fine linear scar whereas a transverse incision is likely to result in an irregular distribution of local tensions which lead to an unsightly scar. Little work has been published on the *in-vivo* levels of biaxial stress within the skin throughout the human body, although a number of attempts have been made to measure the in-plane stresses and resting tension in the skin. However, variations in skin thickness and the orientation in which measurements are taken make it difficult to obtain accurate *in-vivo* estimates of in-plane skin tension either at rest or while stretched. Early work using a suction cup device [7, 8] showed that the ratio of tension parallel and perpendicular to Langer lines on the upper backs of adult males was in the range of 5:1. Due to difficulties in measuring *in-vivo* skin thickness, they reported resting tension values in the region of 4-30 N/m on the upper back of adult males: this suggests that the range of resting stress is in the range of 1-30kPa (assuming skin thickness to be between 1-4 mm). More recent work on the human forearm also suggests that the in-plane resting tension in skin corresponds to stresses in the order of 10kPa [9] or, including effects of subject variability and measurement reproducibility, in the region of 7-20kPa [10].

Tests on the mechanical properties of human skin *in-vitro* [11-13] report that tensile strength ranges from 5-30 MPa. The mean value was 21 MPa at age 8, which decreased to 17 MPa at age 95. The strain at rupture varied from 35%-115% depending on the age and area of the body. The mean rupture strains declined approximately linearly from a maximum of 75% at birth to a minimum of 60% at age 95. The modulus of elasticity ranged from 15-150 MPa with the mean values being greatest at 70 MPa at age 11 and at least 60 MPa also at age 95. While these average values only vary by up to ±12%, the total range of values varies by up to one order of magnitude. This large variability in the mechanical properties of skin is due to differences in age, gender, location and orientation from specimen to specimen. A general review [10] of various measurements of *in-vivo* skin properties revealed that Young's modulus varied by more than three orders of magnitude (from 0.02-57MPa), depending on the test methodology that had been used. The same authors used a suction chamber and an ultrasound device to measure *in-vivo* vertical displacements of the skin surface and of the skin thickness. This information was subsequently used in a simple elastic model to infer the *in-vivo* Young's modulus of skin from the forearm of healthy males aged between 20-30 as being in the range of 129±88kPa.

Knives that are commonly used in homicide and attack situations include those that are most widely available amongst the general population, such as household knives. Figure 1 illustrates the principal parts of a common knife: it is the geometry and shape of the point and tip of a knife that are of most relevance in stabbing [14]. In some knife designs, such as a bread knife, the point is less pronounced and can actually be rounded or curved. The blade edge can be straight or serrated and can be present on one or both sides of the knife. If the edge of a blade is serrated, the serrations generally are not present in the immediate vicinity of the point: their aim is to provide prolonged sharpness to the blade edge during reciprocating cutting and not to control initial penetration by the point of a blade.

It is possible to differentiate between the kinematics of knife stabbing attacks [15]. Overarm attacks involve an overarm motion in which, when gripped, the blade of the knife emanates from the ulnar aspect of the hand. Underarm attacks, on the other hand, involve an underarm movement whereby the blade emanates from the radial aspect of the hand. In overarm attacks, the knife speed upon entry is approximately 10 m/s whilst for underarm attacks it is closer to 7 m/s.  Injuries to the human head including scalp laceration and skull fracture, by contrast, can occur from impact against blunt objects at linear velocities in the order of 4 m/s [16-18] whereas impacts against sharp objects at even slower speeds can also lead to soft tissue cranial injuries [19, 20].  It is reasonable to assume that slow or walk-on stab speeds would occur at velocities in the order of 1.0 m/s, i.e., speeds comparable to typical walking speeds.  However, there is a paucity of published literature available on the mechanics of knife stabbings at walk-on speeds and at quasi-static speeds (i.e., approaching zero velocity).  It is unknown whether the forces associated with fast stab speeds are likely to be significantly different from a similar stab wound at slow speeds: inertial effects would suggest that less force might need to be applied to cause a similar stab wound at fast speeds than at slow speeds.  Viscoelastic materials will exhibit rate dependent effects and typically behave in a relatively brittle manner at fast loading rates and a more ductile manner at slow loading rates: material stiffness, failure strengths and failure strains will depend on stabbing speeds.  Experiments which could be done at quasi-static speeds would remove the effects of inertia, momentum and speed on the force required to penetrate skin and would usefully serve as a baseline against which the effects of inertia and viscoelasticity could be compared.

The geometry and sharpness of a knife greatly affect the ability of the knife to penetrate skin.  Chadwick et al., [21, 22] investigated the biomechanics of knife stab attacks using an instrumented knife and found that mechanical test methods that simulate the kinematics of a knife can accurately ascertain the values of axial force and energy. They found that non-axial forces and torques which occur in stabbing incidents are not easy to reproduce accurately using mechanical methods. There has been little research into the effect of knife angle on penetration forces or the effect of the anisotropic nature of skin on penetration forces.  While there is no accepted definition of knife sharpness, it is clearly related to the geometry of the knife and it is assumed that a sharp knife requires less force to push through soft tissue than a blunt one.  Recent research [23-25] derived a numerical index for the sharpness of a blade edge which does not explicitly depend on the blade geometry or the stiffness or strength of the tissue being cut.  This approach appears to provide quantitative results that are more consistent than previous attempts to characterise the cutting quality of a knife in terms of either maximum cutting force [26-28], the radius of the cutting edge [29-31], or the power [32] or energy [33] used during cutting.  The definition of a dimensionless index for sharpness usefully permitted different combinations of blades and skin simulants to be compared directly in a quantifiable manner.  However, it is the point of the blade that first contacts and penetrates the skin in stabbing incidents and once the skin has been penetrated no increased force is required [34] to penetrate underlying soft tissue. This suggests that the geometry and condition of the point of the knife are more important than those of the blade edge when determining the penetrative ability of a particular knife.

The biaxial measurement device that was developed is described in detail in the following section, along with the design features that make it particularly suitable for understanding the mechanics of skin stabbing.  After confirming the reliability and repeatability of the measurement device, its use in characterising the quantitative differences between blade types under different kinematic conditions is illustrated.  The data reported in this paper concentrates on the kinematics of quasi-static stabbing speeds in order to address the deficit of published literature dealing with slow and walk-on stabbing speeds and to provide a basis against which dynamic effects can subsequently be quantified. While some useful attempts have been made by others to experimentally quantify the forces involved in stabbings [28, 35, 36], these generally have relied on using animal carcasses or excised cadaver tissue. Such an approach always has significant experimental scatter as a result of the inherent anisotropy and variability of biological tissue, assuming, of course, that a statistically significant number of test specimens have been subjected to identical tests. The present investigation clearly demonstrates this large scatter by testing animal skin specimens but by using synthetic skin surrogates that are isotropic and homogeneous, it has been possible to successfully identify the influence of other variables on the mechanics of stabbing. Other experimental investigations that have used instrumented devices to quantify the forces [34-37] have tended to incorporate a single linear force transducer in the handle of a knife to measure forces in the direction of stabbing. The disadvantage of such an approach is that it is practically impossible for an individual to avoid applying small additional components of force in other directions and without any twisting motion when using such a device and such extraneous components of force are not recorded. The present apparatus removes all of this uncertainty and sources of extraneous forces in stabbing experiments by fixing the knife blades to a screw driven uniaxial testing machine and by clamping the skin specimens to an independently controlled set of biaxially aligned load cells.

The motivation behind this work has not been to develop a device that could be used to identify or eliminate a particular knife as the instrument producing a specific stab wound. Rather, the aim has been to develop a procedure that will systematically permit each and every variable to be analysed separately so as to quantify its importance on stab penetration forces. It is hoped that this will assist forensic science to establish the relative significance of each of the multifactorial variables associated with a stabbing and to quantify the extent to which penetration force is influenced by changes in each variable.

**MATERIALS AND METHODS**

Prior to developing the measurement device, consideration was given to the necessary levels of design functionality for the device. Cruciform test specimens, which are conveniently obtained by using a cruciform cutting die of arm width 10mm and arm length 120mm (ODC Tooling & Molds, Ontario, Canada), would need to be held in a varyingly controlled manner to reproduce the range of unequal biaxial stresses that exist *in-vivo* within skin. In none of the experiments that are described in the following section, did the levels of biaxial stress induce any wrinkles on the skin [38]. Given the range of skin thicknesses over the human body, and the different compositions of underlying tissue including fat and muscle, it was also considered appropriate that the device should be capable of accommodating stratified specimens of varying thicknesses and viscoelastic stiffness characteristics. More crucially, since it is skin rather than underlying tissue that is most resistant to penetration, it was considered essential to be able to test specimens of skin in the absence of any underlying tissue. In order to develop a device that would consistently and repeatedly provide accurate measurements of the penetration force characteristics, and be shown to do so, the inherent variability between successive skin samples means that it is not appropriate to use actual biological skin samples. Rather, it is considered more appropriate to use a homogeneous synthetic material which has similar stress-strain characteristics to that of skin tissue. The use of such a skin simulant will ensure that variations in the measured properties during knife penetration tests are solely due to the variables of knife geometry and stab mechanics and not to variations between successive tissue specimens. A material which is suitable and convenient is polyurethane of Shore hardness 40A [3, 23, 24]: it has a J-shaped stress-strain curve with a strength of approximately 17 MPa and a final stiffness of 22 MPa, as measured and illustrated in Fig. 2: this is qualitatively similar to biological skin. Moreover, it is isotropic, as attested by the superposition of the longitudinal and transverse test curves. Ballistic soap and a compliant foam were used [39] to simulate the influence of cartilage and subcutaneous fat underneath the skin.

Fig. 3 shows the physical test setup with the polyurethane skin simulant specimen held under biaxial stress by two opposing pairs of clamps. Each pair can be moved independently from the other, in order to create a state of induced anisotropy. The clamping mechanism for the skin specimens was designed so as to permit the tension or stress in the pair of cruciform arms to be controlled separately by means of two independent load cells (Interface Force model SML10; www.interfaceforce.co.uk), the capacity of which was approximately 45N (10 lbf) with an accuracy of 0.05%. The blade of the knives were held in a separate clamp, also shown in Fig. 3, which was in turn attached to a load cell (not shown) of a uniaxial testing machine (Tinius Olson). This load cell and the knife blade were aligned perpendicularly to the plane of the clamped skin specimens and the testing machine was operated in a displacement mode of control. This meant that, as the machine actuator forced the blade to indent and then penetrate a skin stimulant specimen, the actuator velocity corresponded directly to the out of plane (i.e., downwards) displacement of the specimen. It is this displacement and the corresponding force as applied by the load cell of the testing machine that are analysed in the following sections. An amplifier (Interface model number SY038) was used to amplify and transmit the output from the three load cells at a sampling rate of 21 Hz via a National Instruments data acquisition card (USB 6008) to a PC for subsequent analysis.

Previous experience [3, 40, 41] suggested that the knife and skin samples would best be treated as separate components. This offers the advantage of incorporating a knife holder in the load cell of a uniaxial testing frame instead of either instrumenting a range of knife handles or of measuring out of plane reaction forces induced by a knife on a skin sample. This approach would also facilitate varying the angle and rate at which the blade penetrates the skin, changing the blades between successive tests, and analysing the effects of blade type. In order to avoid having to accommodate different knife handles, and to ensure the point of a knife was introduced consistently into the centre of the cruciform test specimens, blades were centred in a holder using a key in the main body of the holder and a slot ground into each blade, as illustrated in Fig. 4. An advantage of separating the knife clamp from the specimen clamp was that it was immediately possible to incorporate the two fixtures into different testing frames, such as a screw driven or servohydraulic machine or a drop weight tower, depending on whether slow, moderate or high knife speeds were to be simulated. The baseplate of the specimen clamp was attached to an angled plate (not shown), which could be tilted to arbitrary angles in order to simulate varying knife penetration angles.

Four commonly available household knives were chosen for analysis, namely, cook's knife, carving knife, utility knife and kitchen knife (Kitchen Devils, Fiskars, Bridgend, UK). The composition of these knives is confidential proprietary information but chemical analysis indicates that the material is stainless steel (European grade X30Cr13). A standard procedure was followed for sectioning the blades of the knives, taking care at all times to avoid affecting the point and tip regions of the blades. The tip angle of each knife was bisected and this bisector line was noted on each knife (see dashed line in Fig. 5). A line perpendicular to the bisector line was drawn 50mm from the point and the knives were then cut and notched along this perpendicular line using a cut-off saw with a blade suitable for hardened steel. This conveniently allowed a reference orientation, i.e., 0º, to be defined in which the bisector line is normal to the plane of the skin specimens and positive or negative angled orientations refer to the specimen being tilted towards the spine or edge of the blade, respectively. In other words, positive angles are obtuse with respect to the angle between the knife edge and the specimen whereas negative angles are acute.

The effect of several variables including the knife speed, the tension applied to the specimen, the angle of penetration and the specimen thickness were studied for each of the four knives under the range of conditions detailed in Table 1. In particular, note that there was one equi-biaxial state of tension, leading to isotropic loading in the plane of the skin, and two unequal biaxial states of pre-tension, leading to states of stress-induced anisotropy in the skin. The blades of the four knife types are shown in Fig. 5.

At least three virgin specimens were tested for each condition. Two knife speeds, which differed by one order of magnitude, were considered: the faster of these approaches speeds that would typically be encountered in walk-on stabbing incidents. These speeds spanned the entire range of the uniaxial testing machine used for testing and facilitated close observation of the interaction between knives and surrogate skin specimens during testing. Future work to be reported upon subsequently will involve an instrumented drop weight tower which will facilitate faster knife speeds. The levels of tension present in the arms of the test specimens were chosen primarily for convenience and to demonstrate the capability of the testing device to control the biaxial tension and not to represent any specific site on the human body. The range of knife penetration angles that were evaluated were chosen to correspond to reasonably realistic stabbing situations, although it must be emphasised that no torsion was associated with these experiments.

**RESULTS**

In order to compare the results between each knife, it was decided to define a set of reference test conditions based on 4mm thick polyurethane held in the absence of any underlying surrogate tissue under an equi-biaxial tension of 10N (i.e., nominal stress of 250kPa) being penetrated by a cook's knife at a speed of 50 mm/min which is oriented at 0º (i.e., perpendicular to the skin simulant). Fig. 6 presents results from five repetitions of these reference test conditions using the same new knife for each test: the scatter in measured penetration force (mean = 23.99N) and out of plane deformation (mean = 27.05mm) is less than ±2%, which illustrates the repeatability of the measurement device. Fig. 7 compares the results of the average of these five tests against two different new cook's knives, all three of which are nominally identical, under the same set of test conditions. It is clear that the behaviour of each knife is consistently identical among its own set of results. The scatter in penetration forces and out of place displacements from three test repetitions with each of these other two cook's knives is also less than ±2%. However, it is strikingly apparent that the behaviour of each individual knife differs from that of the other knives (c.f. also Table 2). The range between the maximum and minimum penetration forces of all these tests was almost 100%. This large variability between nominally identical knives can only be a consequence of manufacturing variability in controlling the profile of the point of each blade.

The influence of knife type upon the penetration characteristics is detailed in Fig. 8, in which it is seen that, of these four knives, it is the cook's knife that requires greatest force, energy and out of plane displacement to penetrate the skin and the utility knife that requires least. In other words, the knife with the 'sharpest' point is the utility knife and the 'most blunt' point is the cook's knife. The difference between the minimum and maximum penetration forces and out of plane displacements is in the order of 100%. This clearly illustrates the fact that one particular type of knife could lead to a stab wound of a given severity more easily than another. However, the fact that the profile of the point of a knife blade is not controlled during manufacturing, suggests that different knives could actually lead to equally severe stab wounds if consideration were only given to the magnitudes of the penetration force and depth of penetration of the skin. The geometric details of the point and tip regions of the knives is confidential design information that is not available from the knife manufacturers. Nevertheless, inspection of the photographs and scanning electron micrographs of the knives (Fig. 5) indicates that these are appreciably different between each knife type. In forensic investigations, of course, the physical dimensions of a blade and the visual dimensions of a stab (width, angle and depth) would assist in establishing knife type and stab injury severity.

The influence of specimen thickness, knife penetration speed, in-plane tension and angle of knife orientation are illustrated for the cook's knife in Figs 9, 10, 11 and 12 respectively. For each result, either the average or all of at least three test repetitions are presented: the results of each set of tests are typically within ±2% in all cases.

The influence of different stiffness of underlying tissue beneath the skin was illustrated by incorporating either ballistic soap or a compliant synthetic foam underneath the test specimens. The results of these tests are illustrated in Figs 13(a) and (b) for two different types of knives. Ballistic soap has stiffness characteristics that are broadly comparable to costal cartilage whereas those of compliant foams and rubbers are closer to the stiffness of subcutaneous fat [39, 41, 42].

The difficulties inherent in using samples of biological skin tissue for quantitative investigation are illustrated in Fig 14(a) and (b), which presents results obtained by using different knives to stab cured calf leather specimens (Charles Owen Ltd., UK). Repeatable results with little scatter are extremely difficult to obtain with such natural materials, regardless of knife type, unlike when using surrogate skin specimens made from synthetic materials, as in Fig. 6. The scatter associated with the energy required to penetrate these biological skin specimens is in the order of ±20%. The scatter associated with the strength of samples of fresh biological tissue, even when taken from identical locations on a body and aligned in the same orientation can be even greater: ±50% is not uncommon [42].

## DISCUSSION

A number of important physical measurements can be made from the test results contained in the figures of the previous section, in order to characterise the stabbing process using quantitative terms. The first of these are the skin stiffness at the onset of the indentation stage and the skin stiffness at the penetration stage of a knife stabbing a skin simulant specimen, as identified in Fig. 6. These stiffness values are defined as $K_0$ and $K_f$ respectively; they are conveniently calculated as the slopes of the initial and final stages of each stabbing experiment and are tangent moduli of the skin stimulant (c.f. dashed lines in Fig. 6 for explanation). It is clear that the final stiffness is always greater than the initial stiffness (steeper slope) and that the initial stiffness varies less from specimen to specimen (c.f. standard deviation data in Table 2). Other direct measurements that can be made in each experiment include the maximum force required for the point of a knife to penetrate a skin simulant, $P_{max}$, and the amount of out of plane displacement associated with the onset of skin penetration, $\delta_{max}$. These direct measurements, along with calculated values for energy, are summarised in Table 2 for the complete set of tests described in Figs 6-14. The energy values are the energies required to penetrate the skin simulants by each of the knives and have been calculated by numerical integration of the area under the respective force-displacement curves: these are related to both $P_{max}$, and $\delta_{max}$.

The test results obtained from using any particular knife in this custom designed stab measurement device are consistently reliable and repeatable. This is visually evident, for example, in Fig. 6 and from the statistical data summarised in Table 2: the scatter associated with the measured values for $K_0$, $K_f$, $P_{max}$, $\delta_{max}$ and energy is generally less than ±5%. Of these five parameters, the greatest scatter is associated with the final stiffness measurements, $K_f$, whilst the least scatter is associated with the penetration force, $P_{max}$, and the out of plane displacement, $\delta_{max}$, typically in the order of ±2%. The use of natural skin samples (c.f. Fig. 14 and Table 2), which are inherently more variable than synthetic skin simulants, leads to significantly less accurate and repeatable results: the corresponding scatter can be expected to be at least ±20%. This suggests that this measurement device can be used successfully to characterise and to quantify the parameters associated with the mechanics of stabbing. It also suggests that a large population set of skin samples would be required if statistically meaningful results were to be obtained from stab experiments on natural skin. A limitation of this work, however, is the fact that the synthetic skin surrogate that has been used is not actually biological skin. Consequently, the results presented in this paper are not directly equivalent to what the stab penetration forces would be for human skin.

When the initial and final stiffness values are considered, it is apparent that $K_f$ is always greater than $K_0$. This is a direct consequence of the mechanics of stabbing: indenting the point of a knife into the skin will increase the in-plane tension and thus the stiffness of the skin. Since $K_0$ is associated with the initial stages of contact between the knife and the skin, i.e., the onset of indentation, it cannot be used to characterise when the skin is likely to be penetrated by a knife. The final stiffness, on the other hand, is more closely related to the penetration stage of a stabbing event than the initial stiffness and it is reasonable to assume that $K_f$ could be used to characterise skin penetration. However, of all five parameters, it is $K_f$ that has the largest amount of scatter. While this may be due, in part, to the lack of quality control when manufacturing the point of a knife, it would be difficult to rely solely on this particular parameter when attempting to predict the occurrence of skin penetration. The other three parameters, namely $P_{max}$, $\delta_{max}$ and energy, are all consistently associated with the skin penetration stage of a stabbing event and consequently, it may be appropriate to characterise the onset of a stabbing injury using a metric based on either of these parameters.

The results presented in Figs 6-14 present the variation of force against displacement in which the independent variable, the out of plane displacement, was the control variable. This was achieved by operating the testing machine in a displacement mode of control (instead of load control) in which the blade moved at a constant rate (50 mm/min or 500 mm/min). This allows a direct one-to-one comparison to be made between the displacement data and the time taken to achieve this depth of out of plane displacement: in other words, if skin penetration occurs at a displacement value of 25mm, this corresponds to a time of 30 seconds for the reference test rate of 50 mm/min (3 seconds for 500 mm/min).

While the broad level of forces required for the four different types of knives to penetrate the skin simulant material was in the order of 20N (ranged from 18-36N), the energy required was approximately 0.25J (ranged from 0.17-0.56J) and the out of plane displacement was in the order of 20mm. These levels of energy are two orders of magnitude below the 43J threshold [14] set for stab-resistant vests, the value of which was set to correspond to the energies an adult male is typically capable of exerting with a knife. These levels of penetration force are comparable to the levels of force that have been reported [34] to penetrate cadaveric human tissue. Consequently, the levels of force or energy that are measured in the present experiments as being associated with penetration of the skin simulant would correspond to nothing more than a "mild" level of force. It is important to remember, however, that these force and energy measurements involve a skin simulant under levels of stress that do not correspond exactly to *in-vivo* conditions of human skin and it is, therefore, not possible to directly equate these simulant results with *in-vivo* human conditions.

The thickness of the skin simulant has a direct influence on all five of the parameters associated with stabbing, as seen clearly in Fig. 9. For all other conditions being equal, a thinner skin will be associated with lower values for initial and final stiffness and penetration force and energy than a thicker skin. However, a thinner skin will be associated with greater out of plane displacements than a thicker skin: this is due to its stiffness being lower. What is also apparent, from both Fig. 9 and Table 2, is that this direct influence of skin thickness does not appear to vary in direct proportion to the relative differences in skin thickness. In other words, the stab measurements that arise from skin that is half as thick as another skin are not half those of the thicker skin.

The speed with which a knife stabs a skin does not have any appreciable effect on the stabbing results for the two speeds tested in this present investigation, as seen in Fig. 10 and Table 2. It must be emphasised, however, that these speeds are quite slow (less than 0.01m/s) and it is probable that differences would be noticed at faster speeds, especially those speeds typically associated with run-on stabbings [15], i.e., in the order of 10m/s. It is noted, in passing, that of all the five parameters, the final stiffness, $K_f$, does appear to increase with knife speed. While the speeds considered in this investigation are quasi-static, these results will usefully serve as a benchmark comparison against future high speed experiments which will establish the influence of inertial effects on stab penetration forces for skin.

When the four different types of knife blades are considered, as seen in Fig. 8 and Table 2, it is clear that the amounts of force required to penetrate the skin differed by about a factor of two. The force required to initiate a stab wound with a utility knife was approximately half that with a cooking knife. The same was also true for the energy required and measured out of plane displacement values. The implication of this result is that correspondingly severe injuries would require appreciably different levels of force if different types of knives were assumed to have been used during an assault. In other words, the wound that would occur from using a "mild" level of force with one knife may correspond to that of a "moderate" level of force with a different knife. This is consistent with the intuitive expectations of how knives of different geometries, as shown in Fig. 5, are likely to behave during a stabbing incident [24, 25]. The knife tip geometries of the utility and carving knives are slightly more acute (i.e., smaller blade angles) than the carving and cook's knives, even though it is not possible to accurately measure a radius at the point of the knives.

While this can be inferred solely from the results of Fig. 8, i.e., the skin simulant being stabbed in the absence of any underlying surrogate tissue, Fig. 13 provides some additional insight to whether this may also be true in the presence of underlying tissue. In both Figs. 13a and 13b the stage at which the skin is penetrated, i.e., at which the stab wound initiates, corresponds to the first local maximum value of force (not a global maximum value). This is true for both knife types that were tested and when either type of underlying tissue was present. While these local maxima values are appreciably lower than the skin penetration force in the absence of any underlying tissue for the kitchen knife (Fig. 13a), they are comparable for the cook's knife (Fig. 13b). Also, in all of the tests in which either type of underlying tissue was present, the force values drop after this local maximum value has been reached. They then reach a local minimum value, and begin to rise again to a second local maximum value (global maximum in Fig. 13a; tests of Fig. 13b were stopped before similar global maxima were reached). The first inference to be made from these tests where either soft or firm underlying tissue has been present is that a stab may initiate or skin can be punctured just as easily in either case: similar levels of force may only be required. The associated depth of a wound will depend on the stiffness and thickness of underlying tissue: the least out of plane displacements are associated with the stiffest underlying tissue. Tests that purposely avoid using any surrogate underlying tissue can provide useful insight into the mechanics of a knife stab: the 'free' data of Fig. 13 are broadly similar to the 'foam' and 'soap' results and differences can be explained in terms of the stiffness and thickness of surrogate underlying tissues. The initial drop in force after reaching local maxima are due to elastic strain energy being released and in-plane skin tension reducing after the point of a knife has penetrated the skin. The subsequent rise in force from local minimum values is due to the point, tip and edge of a knife penetrating the underlying tissue: increasingly larger cross-sections of a knife are entering into the underlying tissue. It is important to realise that this does not mean that the force required to penetrate underlying tissue is greater than that required to puncture skin.

It has been shown in Fig. 7 that the manufacturing process for these knives does not accurately control the machined finish or precise geometric profile of the point of the knife blades. This has important implications for forensic analyses in which knowledge of the particular type of knife that was used to cause a given stab wound is correlated against the level of force that may have been applied. Variability in a given type of knife may well eclipse variance from other factors and forensic analyses should take careful account of knife blade conditions. Since it is probable that different knife manufacturers use similar processes to manufacture their knives, none of which are purposely designed specifically for stabbing, it is likely that this degree of variability is common to all household knives and not solely limited to those manufactured by Fiskars.

Fig. 11 presents a useful insight into the influence of an unequal biaxial state of pre-tension in the skin simulant. The reference test conditions of Fig. 6 considered an equi-biaxial state of tension (10N:10N) in the opposing arms of the cruciform skin samples. Fig. 11, on the other hand, has a corresponding state of tension of 25N:5N. The ratio of 5:1 was purposely chosen to correspond to the measured ratio of *in-vivo* tension parallel and perpendicular to Langer lines on the human body [7, 8]. A corresponding ratio of 50:10 was not chosen since the maximum rating of the load cells was in the order of 45N (10 lbf) whilst the ratio of 25:5 was chosen in favour of 10:2 since the larger levels of absolute tension would emphasise any observed differences. The two sets of results given in Fig. 11 and summarised in Table 2 show a marked difference depending on whether the greatest level of skin tension is parallel to or perpendicular to the plane of the blade as the knife point penetrates the skin simulant. This difference is in the order of 20% for the levels of force, out of plane displacement and energy. The measured values for these parameters are all greater when the 25N skin tension level is perpendicular to the plane of the blade and the 5N is parallel to the plane of the blade. The corresponding values are all lower when the greater level of skin tension is parallel to the blade. This is directly consistent with the observed behaviour when cutting biological skin: less force is required when cutting parallel to the Langer lines and a greater level of force is required when cutting perpendicular. In other words, if the orientation of the Langer lines in biological skin correspond to the direction of greatest tension (25N) in the skin simulant, it is easier to penetrate and cut the skin when the plane of the knife blade is parallel to this direction but greater force is required if the plane of the blade is perpendicular to this direction.

The same cook's knife was used for these unequal biaxial tension tests (CK1) as the 'CK1' data of Fig. 7. When the levels of force, out of plane displacement and energy are compared for the reference conditions (10N:10N) for this blade and for these unequal biaxial tension tests, some additional interesting observations can be made. Consider the results associated with different levels of tension parallel to the plane of the blade. Increasing levels of tension (5N Vs 10N Vs 25N) progressively lead to reduced amounts of force and energy being necessary to penetrate the skin: it becomes progressively easier to penetrate and cut the skin as the tension parallel to the blade increases. Consider also the trend when the level of tension perpendicular to the plane of the blade increases (5N Vs 10N Vs 25N): progressively more force and energy are required to penetrate the skin.

This information can be interpreted with respect to the *in-vivo* levels of skin tension that exist within a human. These levels of tension (i.e., 5N, 10N and 25N) equate to stresses in the range of 125-625 kPa. Literature discussed earlier [7-10] suggests that the at-rest skin stresses within the human body are in the region of 1-30 kPa, i.e., approximately one tenth of those tested in the present experiments. However, during most assaults and during many surgical procedures the human skin is tautened and tensed: the *in-vivo* skin stresses under such conditions will be greater than at-rest levels, although not necessarily as high as the 125 kPa of the present experiments. Most fatal wounds received in assaults are to the trunk of the human torso [1] and skin tension around an assault wound in this region is likely to be relatively close to at-rest levels of tension. Consequently, it is important to realise that the results of the present experiments on a skin simulant material were at levels of stress that exceeded those of *in-vivo* assault conditions. Therefore, it is not possible to directly equate these simulant results with *in-vivo* human conditions. Certainly, the corresponding levels of penetration force and energy on simulant skin materials would be greater than measured in the present experiments if the tension/stress levels were closer to at-rest in-vivo levels.

The angle at which the point of the blade entered the skin simulant specimens did have some influence on the measured values of penetration forces, out of plane displacements and energies. However, this variation was modest for the range of angles that were considered in this present investigation, namely -15º to +15º with respect to the defined 0º reference axis. The results in the range –5º to +5º were all reasonably close to the reference data of 0º whilst the forces and energies associated with larger positive angles (+10º and +15º) were approximately 10% less than with negative angles (-10º and -15º, respectively). It is possible that more pronounced differences would have been apparent if a different 0º reference angle had been defined. An alternative datum line instead of the bisector line mentioned earlier and indicated in Fig. 5 (dashed line) could have been defined by taking a line from the point of the knife to the butt of the knife handle (c.f. solid line in Fig. 5). The differences between the bisector lines used in this present investigation and such an alternative definition of a datum line are +2.5º for the carving knife, +8º for the cook's knife, +10º for the kitchen knife and +15º for the utility knife. In other words, the 0º reference results of the present investigation should be identical to those of +2.5º for the carving knife etc. if the alternative definition of a datum line had been used.

Recent innovative work by McCarthy et al. [24] describes the development of a blade sharpness index. This is a dimensionless numerical value that conveniently allows the sharpness of different types of blades to be compared in a quantitative manner during the cutting of different types of materials. The sharpness index normalises the energy required to initiate a cut (e.g., puncture) in a material with a given knife against the physical characteristics of the particular material (critical strain energy release rate and thickness) and the material displacement at the onset of cutting. While it would potentially be useful to develop an analogous blade 'stab' index, this is not possible solely from the results of the present experiments since the fracture mechanics characteristics ahead of and at the point of a blade during the stabbing process are unknown. Future work could usefully attempt to define the relative contributions of Modes I, II and III fracture energy [24] (related to the local tensile and shear stress states within a material immediately ahead of an advancing blade or crack) during a stab event with a particular knife and consequently deduce a blade stab index which would allow direct comparisons to be made between the points of different types of knives.

**CONCLUSIONS**

A biaxial test device has been developed that can reliably and repeatably quantify the characteristics associated with knife stab penetrations of synthetic skin simulants. The device can easily permit the influence of different knife tip configurations to be identified and their effect of penetration forces to be quantified. This offers significant advantages over instrumented knives where forces are measured by transducers that are embedded in the knife handle: this measurement device offers greater control and variability, a real-time display of results and the ability to simulate different skin tensions replicating different body sites. The measurement device allows for independent control of key variables that affect the mechanics of stabbing, namely, the skin thickness, the blade type, the levels of biaxial tension within the skin, the presence of various substrate materials, the angle of attack of the knife and the test speed. Measured parameters that most accurately characterise the occurrence of skin penetration (i.e., the stab) are seen to be the force and energy applied by the knife, and the out of plane displacement of the skin. This skin displacement is directly related to the elapsed time after the knife first comes in contact with the skin. The scatter associated with these test parameters was in the order of ±2%. Scatter associated with specimens of biological skin displayed an order of magnitude more scatter (±20%).

Four different commonly available household knives (cook's, utility, carving and kitchen knives) were tested. The utility knife required the least amount of force or energy to penetrate the skin and was associated with the smallest amount of out of plane skin displacement, while the cook's knife required the greatest force, energy and out of plane displacement. While the utility knife was therefore judged to be 'sharpest' and the cook's knife the most 'blunt', the profile of the point of a knife blade was shown to be not directly controlled during manufacturing. The consequences of this are that different types of knives can actually require either equal or different amounts of force to cause comparable stab wounds. It can also mean that the penetration forces associated with nominally identical knives, even virgin knives, can vary by as much as 100%.

The test speeds considered in this investigation (50 mm/min and 500 mm/min) did not influence the test results. However, these speeds are quite slow (less than 0.01m/s) and it is probable that inertial effects would become apparent at faster speeds, especially those speeds typically associated with run-on stabbings, i.e., in the order of 10m/s.

The levels of unequal biaxial tension within the skin had a marked effect on the stab results. Less force and energy were required to puncture the skin when the plane of the blade was parallel to the direction of greater skin tension whereas greater levels were required when the blade was perpendicular. Similarly too, the absolute levels of skin tension had an appreciable effect on the stab results. Increasing levels of tension within the skin parallel to the plane of the blade are associated with progressively lower levels of force and energy being required to penetrate and cut the skin. These findings are intuitively consistent with the observed behaviour when cutting biological skin: less force is required to cut parallel to the Langer lines than perpendicularly and less force is required to cut when the skin is under a greater level of tension. The ratio of unequal biaxial tension that was considered in the present investigation corresponds to what exists in the human body; the actual levels of tension correspond more closely to the tension that would exist within the skin of a human during an assault or undergoing surgical incisions than they would under at-rest body conditions.

It is planned that future publications should address a number of other important factors that are beyond the scope of the present article. These will include the influence of faster knife speeds on the dynamic response of the skin during stabbing, the effect of other biological variables including different thicknesses and stiffnesses of layers of underlying tissue, the correlation between the dimensions of observed tears in clothing and the physical dimensions of a stab wound, and the influence of torsion on the kinematics of stabbing.

**REFERENCES**

[1]   M.T. Cassidy and M. Curtis, Victims of penetrating and incised wounds: A review of 200 cases, in M.D. Gilchrist (Ed.) Impact Biomechanics: From Fundamental Insights to Applications, Springer, The Netherlands, 2005, pp. 405-414.
[2]   J.E. Sanders, B.S. Goldstein and D.F. Leotta, 1995, Skin response to mechanical stress: Adaptation rather than breakdown – A review of the literature. Journal of Rehabilitation Research and Development, 32 (1995) 214-228.
[3]   E. O'Dwyer, Experimental Cutting of Surrogate Soft Biological Tissue with Unserrated Blades. MEngSc thesis, Mechanical Engineering, University College Dublin, 2005.
[4]   G. Wilkes, I. Brown and R. Wildnauer, The biomechanical properties of skin, CRC Critical Reviews in Bioengineering, (1973) 453-495.
[5]   Y. Wu, D. Thalmann and N. Magnenat Thalmann, A plastic visco-elastic model for wrinkles in facial animation and skin ageing, Journal of Visualization and Computer Animation, 6 (1995)195-205.
[6]   K. Langer, On the anatomy and physiology of the skin: 2 – Skin tension. The British Journal of Plastic Surgery, 31 (1978) 93-106.
[7]   H. Alexander and T.H. Cook, Accounting for natural tension in the mechanical testing of human skin. The Journal of Investigative Dermatology, 69 (1977) 310-314.
[8]   T. Cook, H. Alexander and M. Cohen, Experimental method for determining the 2-dimensional mechanical properties of living human skin. Medical and Biological Engineering & Computing, 15 (1977) 381-390.
[9]   J. Emmanuelle, G. Josse, K. Fourad and G. Camille, A new experimental method for measuring skin's natural tension. Skin Research & Technology, in press (2007) **doi:10.1111/j.1600-0846.2007.00259.x**
[10]  S. Diridollou, F. Patat, F. Gens, L. Vaillant, D. Black, J.M. Lagarde, Y. Gall and M. Berson, In vivo model of the mechanical properties of the human skin under suction, Skin Research and Technology, 6 (2000) 214-221.

[11] H. Vogel, Age dependence of mechanical and biochemical properties of human skin, Part I: Stress-strain experiments, skin thickness and biochemical analysis. Bioenginering of Skin, 3 (1987) 67-91.
[12] H. Vogel, Age dependence of mechanical and biochemical properties of human skin, Part II: Hysteresis, relaxation, creep and repeated strain experiments. Bioenginering of Skin, 3 (1987) 141-176.
[13] C. Edwards and R. Marks, Evaluation of biomechanical properties of human skin. Clinics in Dermatology, 13 (1995) 375-380.
[14] J. Croft, PSDB Body Armour Standards for UK Police: Part 3: Knife and Spike Resistance. Police Scientific Development Branch, UK (2003).
[15] S. Miller and M. Jones, Kinematics of four methods of stabbing: A preliminary study. Forensic Science International, 82 (1996) 183-190.
[16] K. O'Riordain, P.M. Thomas, J.P. Phillips and M.D. Gilchrist, Reconstruction of real world head injury accidents resulting from falls using multibody dynamics. Clinical Biomechanics, 18 (2003) 590-600.
[17] M.D. Gilchrist, D. O'Donoghue and T. Horgan, A two dimensional analysis of the biomechanics of frontal and occipital head impact injuries. International Journal of Crashworthiness, 6 (2001) 253-262.
[18] T.J. Horgan and M.D. Gilchrist, The creation of three-dimensional finite element models for simulating head impact biomechanics. International Journal of Crashworthiness, 8 (2003) 353-366.
[19] M.D. Gilchrist, Modelling and accident reconstruction of head impact injuries. Key Engineering Materials, 245-246 (2003) 417-430.
[20] M.D. Gilchrist, Experimental device for simulating traumatic brain injury resulting from linear accelerations. Strain, 40 (2004) 180-192.
[21] E. Chadwick, A. Nicola, J. Lanea and T. Gray, Biomechanics of knife stab attacks. Forensic Science International, 105 (1999) 35-44.
[22] E. Chadwick, A. Nicola, S. Floyd and T. Gray, A telemetry-based device to determine the force displacement behaviour of materials in high impact loading situations. Journal of Biomechanics, 33 (2000) 361-365.
[23] C.T. McCarthy, M. Hussey and M.D. Gilchrist, An investigation into the forces generated when cutting biomaterials with surgical scalpel blades. Key Engineering Materials, 293-294 (2005) 769-776.
[24] C.T. McCarthy, M. Hussey and M.D. Gilchrist, On the sharpness of straight edge blades in cutting soft solids: Part I – indentation experiments. Engineering Fracture Mechanics, 74 (2007) 2205-2224.
[25] C.T. McCarthy, M. Hussey and M.D. Gilchrist, On the sharpness of straight edge blades in cutting soft solids: Part II – Numerical simulations. Engineering Fracture Mechanics, Submitted.
[26] F.H. Watkins, S.D. London, J.G. Neal, J.G. Thacker and R.F. Edlich, Biomechanical performance of cutting edge surgical needles. Journal of Emergency Medicine, 15 (1997) 679-685.
[27] J.G. Thacker, G.T. Rodeheaver, M.A. Towler and R.F. Edlich, Surgical needle sharpness. American Journal of Surgery, 157 (1989) 334-339.
[28] R.W. McGorry, P.C. Dowd and P.G. Dempsey, A technique for field measurement of knife sharpness. Applied Ergonomics, 36 (2005) 635-640.

[29]  T.E. Popowics and M. Fortelius, On the cutting edge: tooth blade sharpness in herbivorous and faunivorous mammals. Annals Zoologica Fennici, 34 (1997) 73-88.

[30]  A.R. Evans, Connecting morphology, function and tooth wear in microchiropterans. Biological Journal of the Linnean Society, 85 (2005) 81-96.

[31]  J.J. Yuan, M. Zhou and S. Dong, Effect of diamond tool sharpness on minimum cutting thickness and cutting surface integrity in ultraprecision machining. Journal of Materials Processing Technology, 62 (1996) 327-330.

[32]  H.J. Huebscher, G.J. Gober and P.K. Lommatzsch, The sharpness of incision instruments in corneal tissue. Ophthalmic Surgery, 20 (1989) 120-123.

[33]  Y. Ueno, M. Asano, H. Nushida, J. Adachi and Y. Tatsuno, An unusual case of suicide by stabbing with a falling weighted dagger. Forensic Science International 101 (1999) 229-236.

[34]  P.T. O'Callaghan, M.D. Jones, D.S. James, S. Leadbeatter, C.A. Holt and L.D.M. Nokes, Dynamics of stab wounds: Force required for penetration of various cadaveric human tissues. Forensic Science International, 104 (1999) 173-178.

[35]  B. Knight, The dynamics of stab wounds. Forensic Science International, 6 (1975) 249-255.

[36]  M.A. Green, Stab wound dynamics – A recording technique for use in medico legal investigations. Journal of the Forensic Science Society, 18 (1978) 161-163.

[37]  S. Jones, L. Nokes and S. Leadbeatter, The mechanics of stab wounding. Forensic Science International, 67 (1994) 59-63.

[38]  M. Destrade, M.D. Gilchrist and G. Saccomandi, Surface instability of skin. Proceedings of EUROMECH Colloquium 481 *Recent Advances in the Theory and Application of Surface and Edge Waves*. Keele University, June 11-13 (2007) 17-19.

[39]  J. Jussila, A. Leppäniemi, M. Paronen and E. Kulomäki, Ballistic skin simulant. Forensic Science International, 150 (2005) 63-71.

[40]  E. Kelly, Penetration and Steady-state Cutting of Viscoelastic Materials. BE thesis, Mechanical Engineering, University College Dublin, 2006.

[41]  M. Ryan, Kinematics of Blade Penetration of Surrogate Skin: A Bioengineering Approach, MSc thesis, University of Strathclyde, 2006.

[42]  J. Ankersen, A.E. Birkbeck, R.D. Thomson and P. Vanezis, Puncture resistance and tensile strength of skin stimulants. Proceedings of the Institution of Mechanical Engineers, Part H, 213 (1999) 493-501.

| Speed (mm/min) | Biaxial Tension (N:N) (parallel:perpendicular) | Penetration Angle (±º) | Specimen Thickness (mm) |
|---|---|---|---|
| 50; 500 | 10:10; 25:5; 5:25 | 1; 2; 3; 4; 5; 10; 15 | 2; 4 |

*Table 1: Test matrix for the various knives (carving, cook's, utility and kitchen). Note that the unequal biaxial tension is conveniently expressed as the levels parallel to the plane of the blade to those perpendicular to the blade. The reference conditions were selected as a knife speed of 50 mm/min, equi-biaxial skin tension of 10N:10N, a penetration angle of 0º and a skin thickness of 4 mm.*

| Figure | Specimen | $K_0$ [N/mm] | $K_f$ [N/mm] | $P_{max}$ [N] | $\delta_{max}$ [mm] | Energy [mJ] |
|---|---|---|---|---|---|---|
| 6 | a | 0.52 | 0.95 | 23.55 | 26.20 | 273.79 |
|   | b | 0.49 | 0.90 | 24.20 | 27.52 | 297.21 |
|   | c | 0.50 | 1.02 | 24.23 | 27.12 | 291.48 |
|   | d | 0.50 | 0.98 | 23.58 | 27.12 | 284.04 |
|   | e | 0.49 | 1.03 | 24.38 | 27.28 | 295.73 |
|   | *Average* | *0.50* | *0.98* | *23.99* | *27.05* | *288.45* |
|   | Std Dev | 0.01 | 0.06 | 0.39 | 0.50 | 9.66 |
| 7 | CK 1a | 0.47 | 0.93 | 17.20 | 22.30 | 176.60 |
|   | CK 1b | 0.47 | 0.96 | 17.56 | 22.50 | 181.77 |
|   | CK 1c | 0.47 | 0.94 | 17.60 | 23.00 | 185.64 |
|   | Average | 0.47 | 0.94 | 17.45 | 22.60 | 181.34 |
|   | Std Dev | 0.00 | 0.02 | 0.22 | 0.36 | 4.54 |
|   | CK 2a | 0.49 | 1.27 | 34.68 | 35.80 | 538.39 |
|   | CK 2b | 0.50 | 0.97 | 33.76 | 35.08 | 513.27 |
|   | CK 2c | 0.49 | 1.41 | 34.24 | 34.92 | 507.73 |
|   | Average | 0.49 | 1.22 | 34.23 | 35.27 | 519.80 |
|   | Std Dev | 0.01 | 0.22 | 0.46 | 0.47 | 16.34 |
| 8 | Carving | 0.44 | 0.78 | 14.51 | 20.00 | 137.54 |
|   | Utility | 0.45 | 0.65 | 13.14 | 18.10 | 114.17 |
|   | Kitchen | 0.53 | 1.21 | 20.23 | 22.80 | 208.54 |
| 9 | 2mm | 0.29 | 0.73 | 14.03 | 27.28 | 165.19 |
|   | 2mm | 0.27 | 0.76 | 15.06 | 29.00 | 187.37 |
|   | 2mm | 0.28 | 0.76 | 14.63 | 28.20 | 177.72 |
|   | Average | 0.28 | 0.75 | 14.57 | 28.16 | 176.76 |
|   | Std Dev | 0.01 | 0.01 | 0.52 | 0.86 | 11.12 |
| 10 | 500a | 0.50 | 1.43 | 23.75 | 26.52 | 273.16 |
|   | 500b | 0.55 | 1.15 | 23.90 | 26.52 | 278.23 |
|   | 500c | 0.51 | 1.29 | 23.93 | 26.52 | 276.47 |
|   | Average | 0.52 | 1.29 | 23.86 | 26.52 | 275.95 |
|   | Std Dev | 0.03 | 0.14 | 0.09 | 0.00 | 2.57 |
| 11 | 5-25a | 0.54 | 0.89 | 15.84 | 19.92 | 150.63 |
|   | 5-25b | 0.49 | 0.93 | 15.58 | 19.80 | 146.20 |
|   | 5-25c | 0.53 | 0.80 | 15.06 | 19.45 | 141.23 |
|   | Average | 0.52 | 0.87 | 15.49 | 19.72 | 146.02 |
|   | Std Dev | 0.03 | 0.07 | 0.40 | 0.24 | 4.70 |
|   | 25-5a | 0.52 | 0.88 | 18.28 | 22.00 | 188.04 |
|   | 25-5b | 0.52 | 0.93 | 18.20 | 21.60 | 184.88 |
|   | 25-5c | 0.53 | 0.95 | 18.50 | 22.00 | 189.40 |
|   | Average | 0.52 | 0.92 | 18.33 | 21.87 | 187.44 |
|   | Std Dev | 0.00 | 0.04 | 0.16 | 0.23 | 2.32 |

*Table 2: Summary results from knife stabbing experiments. Data shown in italics correspond to 'Reference' conditions (Figure 6) that are used for comparison in Figs 7, 8, 9, 10 and 12.*

| | | | | | |
|---|---|---|---|---|---|
| **12a** | 1 | 0.49 | 1.10 | 23.40 | 26.88 | 277.36 |
| | 2 | 0.47 | 1.09 | 23.19 | 27.00 | 275.64 |
| | 3 | 0.50 | 1.09 | 23.55 | 27.08 | 282.28 |
| | 4 | 0.50 | 1.14 | 23.59 | 27.00 | 281.00 |
| | 5 | 0.40 | 1.01 | 23.33 | 26.88 | 275.75 |
| | 10 | 0.50 | 1.03 | 22.33 | 25.92 | 263.22 |
| | 15 | 0.51 | 0.99 | 24.33 | 27.00 | 296.83 |
| **12b** | -1 | 0.49 | 1.10 | 23.40 | 23.40 | 202.93 |
| | -2 | 0.47 | 1.09 | 23.19 | 27.00 | 275.64 |
| | -3 | 0.50 | 1.09 | 23.55 | 27.08 | 282.28 |
| | -4 | 0.50 | 1.14 | 23.59 | 27.00 | 281.00 |
| | -5 | 0.50 | 1.12 | 23.86 | 27.20 | 285.24 |
| | -10 | 0.51 | 1.11 | 24.80 | 27.20 | 299.16 |
| | -15 | 0.53 | 1.33 | 27.18 | 27.00 | 319.45 |
| **13a** **K Knife** | Free | 0.23 | 0.65 | 22.98 | 27.20 | 275.25 |
| | Foam a | 0.59 | 1.19 | 10.82 | 11.80 | 58.56 |
| | Foam b | 0.60 | 1.13 | 10.74 | 11.80 | 58.19 |
| | Foam c | 0.55 | 1.06 | 9.99 | 11.30 | 52.52 |
| | Average | 0.58 | 1.12 | 10.52 | 11.63 | 56.42 |
| | Std Dev | 0.03 | 0.06 | 0.46 | 0.29 | 3.39 |
| | Soap a | 0.57 | 6.05 | 21.08 | 5.38 | 42.97 |
| | Soap b | 0.44 | 5.31 | 14.25 | 5.12 | 37.85 |
| | Soap c | 1.00 | 5.38 | 13.15 | 4.20 | 21.83 |
| | Average | 0.67 | 5.58 | 16.16 | 4.90 | 34.22 |
| | Std Dev | 0.29 | 0.41 | 4.30 | 0.62 | 11.03 |
| **13b** **CK 1** | Free | 0.47 | 0.94 | 17.45 | 22.60 | 181.34 |
| | Foam a | 0.67 | 1.06 | 18.20 | 18.80 | 184.02 |
| | Foam b | 0.65 | 1.05 | 18.58 | 19.00 | 200.24 |
| | Foam c | 0.62 | 0.99 | 18.48 | 19.00 | 187.26 |
| | Average | 0.64 | 1.03 | 18.42 | 18.93 | 190.51 |
| | Std Dev | 0.03 | 0.04 | 0.20 | 0.12 | 8.58 |
| | Soap a | 0.56 | 4.16 | 22.38 | 7.31 | 64.71 |
| | Soap b | 0.49 | 3.92 | 18.32 | 6.63 | 48.38 |
| | Soap c | 0.47 | 3.88 | 17.92 | 6.54 | 47.32 |
| | Average | 0.51 | 3.98 | 19.54 | 6.83 | 53.47 |
| | Std Dev | 0.05 | 0.15 | 2.46 | 0.42 | 9.75 |
| **14a** | Leather a | 0.22 | 1.38 | 11.58 | 15.70 | 70.00 |
| | Leather b | 0.09 | 2.25 | 12.34 | 15.40 | 57.05 |
| | Leather c | 0.08 | 2.15 | 11.36 | 13.70 | 64.53 |
| | Leather d | 0.13 | 1.40 | 8.96 | 13.90 | 42.12 |
| | Average | 0.13 | 1.79 | 11.06 | 14.68 | 58.43 |
| | Std Dev | 0.06 | 0.47 | 1.46 | 1.02 | 12.10 |
| **14b** | Leather a | 0.39 | 0.55 | 3.82 | 5.65 | 11.04 |
| | Leather b | 0.44 | 0.87 | 6.48 | 8.36 | 26.06 |
| | Leather c | 0.44 | 0.78 | 6.89 | 8.03 | 26.67 |
| | Leather d | 0.53 | 0.61 | 6.90 | 8.00 | 27.10 |
| | Average | 0.45 | 0.70 | 6.02 | 7.51 | 22.72 |
| | Std Dev | 0.06 | 0.15 | 1.48 | 1.25 | 7.80 |

*Table 2 continued: Summary results from knife stabbing experiments. Data shown in italics correspond to 'Reference' conditions (Figure 6) that are used for comparison in Figs 7, 8, 9, 10 and 12.*

FIGURE CAPTIONS

Figure 1: Constituent parts of typical knife (Kitchen Devil Cook's Knife) include point (A), tip region (B), cutting edge (C) and spine (D) of blade, and butt of handle (E).

Figure 2: True stress-stretch ratio data for longitudinal and transverse tests on for skin simulant, polyurethane with Shore hardness 40A.

Figure 3: Physical test rig for stab force measurements under controlled biaxial stress state. The simulant skin specimen is cruciform in shape and clamped by two pairs of opposing clamps with load cells (visible behind the top left and top right clamps). The blade is clamped to the load cell (not shown) of a uniaxial testing machine and is perpendicular to the plane of the skin specimen. Forces and displacements recorded by the load cell and actuator to which the blade is attached are analysed to characterise stab penetrations. The simulant material for underlying tissue is removed for clarity.

Figure 4: Schematic of test arrangement showing (top) clamping of blade and (bottom) adjustable stage for clamping cruciform surrogate skin sample under biaxial tension with two pairs of clamps with load cells (seen immediately behind top right and bottom right clamps).

Figure 5: Sectioned blades with locator notches from examples of each of the four knives examined, namely, cook's knife, kitchen knife, carving knife and utility knife (left to right; top photo) and corresponding scanning electron micrograph (SEM; bar markers correspond to 100µm) images of blade tip and point regions (bottom). The knives in the top photograph are perpendicular to plane of skin specimens (i.e., 0º), as indicated by dashed bisector line. The corresponding solid lines define axes from the point to the butt of the knife. The angle between the solid line and dashed line is 8º, 10º, 2.5º and 15º for the cook's, kitchen, carving and utility knives, respectively. The notch in the spine of the utility knife was machined to locate the blade against the screw in the blade clamp (c.f. Fig. 4).

Figure 6: Cook's knife stab test results from five experiments under 'reference' conditions, i.e., 4mm thick skin with no underlying tissue and equi-biaxial tension of 10N in both arms of the cruciform polyurethane skin simulants. Stab speed was 50mm/min and blade was aligned perpendicular to skin (i.e., angle = 0º). The dashed lines illustrate how initial and final tangent moduli for stiffnesses $K_0$ and $K_f$ respectively are calculated.

Figure 7: Cook's knife stab test results under reference conditions: 4mm thick skin specimens without underlying tissue under equi-biaxial tension of 10N at50mm/min at 0º. The average of the reference data of Fig. 6 is shown by the single heavy line (labelled 'CK Ref'). Data from two sets of at least three tests using other new and nominally identical cook's knives are indicated by 'CK 1' and 'CK 2'.

Figure 8: Stab test results for all knives under reference conditions, i.e., 4mm thick skin specimens, 0º angles of knife orientation, no underlying tissue, equi-biaxial tension of 10N; knife speed of 50mm/min. The cook's knife (average of reference data in Fig. 6) is shown by heavy line whereas the utility, carving and kitchen knife curves are the average of at least three test repetitions each.

Figure 9: Stab test results for the same cook's knife penetrating 4mm (average of reference data in Fig. 6 shown above by single heavy line) and 2mm (light lines represent four different tests) thick skin specimens. No underlying tissue was used; equi-biaxial tension of 10N was applied; knife speed and orientation were 50mm/min and 0º, respectively.

Figure 10: Stab test results for the same cook's knife penetrating 4mm thick skin specimens at knife speeds of 50 mm/min (average of reference data in Fig. 6 shown above by single heavy line) and 500 mm/min (the three separate light lines represent three different tests). No underlying tissue was used; equi-biaxial tension of 10N was applied; knife orientation was 0º.

Figure 11: Stab test results for the same cook's knife penetrating 4mm thick skin specimens under different and unequal levels of in-plane biaxial tension; knife speed and orientation were 50mm/min and 0º, respectively. Three tests are under biaxial tension of 5:25N (5N parallel to plane of blade and 25N perpendicular to blade) while three separate tests are under biaxial tension of 25:5N (25N parallel to plane of blade and 5N perpendicular to blade). Note that a new and nominally identical cook's knife ('CK 1' of Fig 7) was used for all of these tests. The star identifies the average $P_{max}$ and $\delta_{max}$ of Fig. 7, which correspond to tension of 10N:10N.

Figure 12: Stab test results for the same cook's knife penetrating 4mm thick skin specimens under different positive angles (1, 2, 3, 4, 5, 10 and 15º) of knife orientation (top; towards spine) and negative angles (-1, -2, -3, -4, -5, -10 and -15º) of knife orientation (bottom; towards edge). No underlying tissue was used; equi-biaxial tension of 10N was applied; knife speed was 50mm/min. The average of reference data in Fig. 6 is shown in both cases by single heavy line.

Figure 13: Stab test results for (top) the same kitchen knife and (bottom) the same cook's knife('CK 1' of Fig. 7 penetrating 4mm thick skin specimens on top of different underlying tissue. Ballistic soap provides stiffer results than compliant foam, while the least stiff results occur in the absence of any underlying tissue. Equi-biaxial tension of 10N was applied; knife speed and orientation were 50mm/min and 0º, respectively.

Figure 14: Stab test results for (top) kitchen knife penetrating prepared specimens of leather and (bottom) cook's knife ('CK 1') penetrating specimens of leather. Equi-biaxial tension of 10N was applied; knife speed and orientation were 50mm/min and 0º, respectively, and no underlying tissue was present in either case. The heavy lines correspond to knives penetrating 4mm thick PU specimens under the same conditions. Note that the calf's leather used for each set of tests (kitchen knife Vs cook's knife) had been cured differently and thus had slightly different mechanical properties.

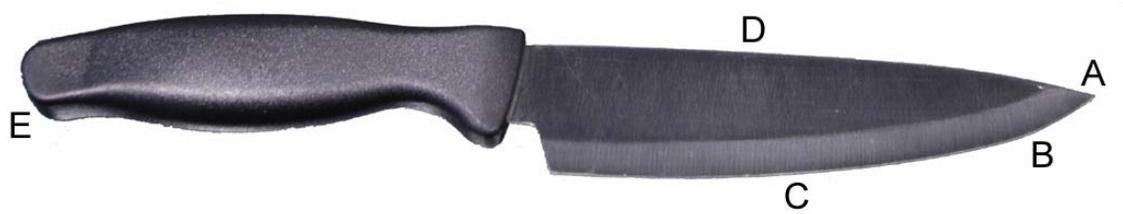

*Figure 1: Constituent parts of typical knife (Kitchen Devil Cook's Knife) include point (A), tip region (B), cutting edge (C) and spine (D) of blade, and butt of handle (E).*

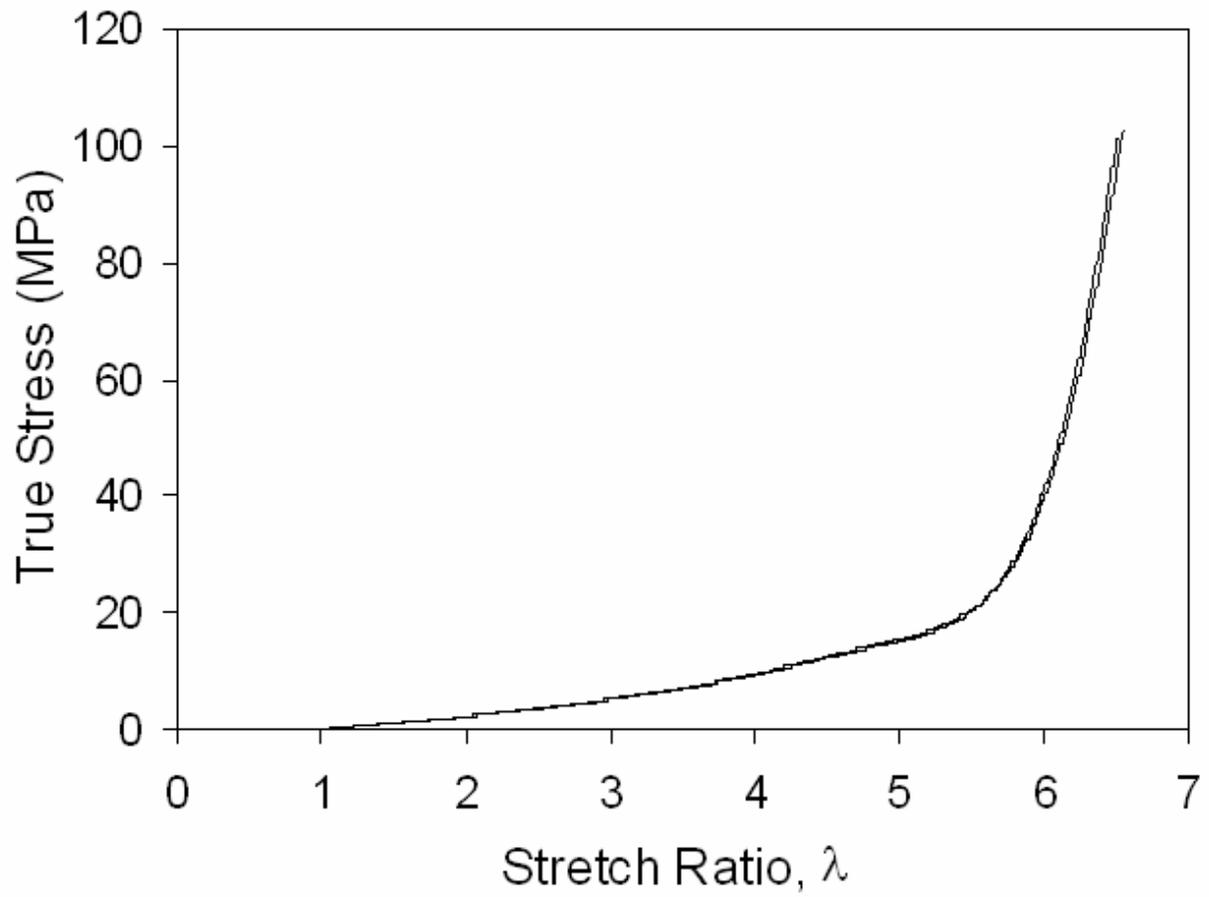

*Figure 2: True stress-stretch ratio data for longitudinal and transverse tests on for skin simulant, polyurethane with Shore hardness 40A.*

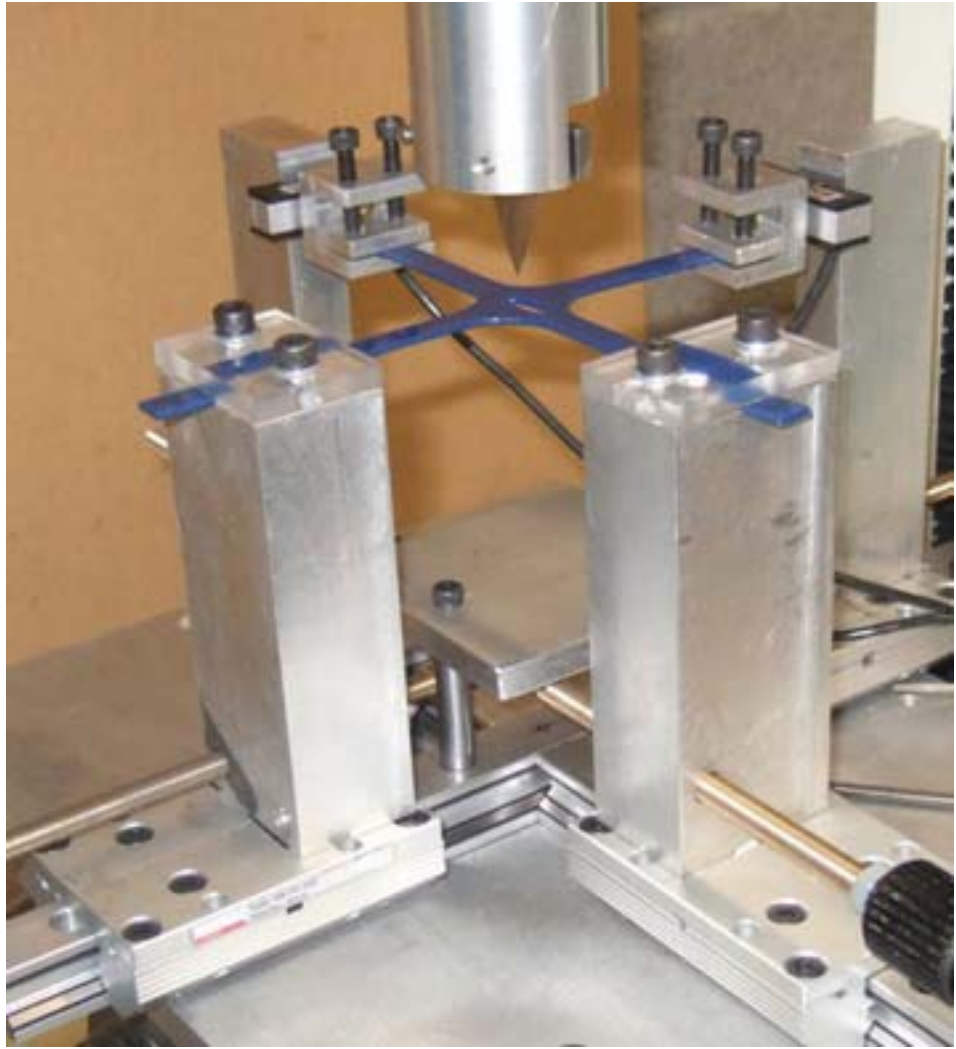

*Figure 3: Physical test rig for stab force measurements under controlled biaxial stress state. The simulant skin specimen is cruciform in shape and clamped by two pairs of opposing clamps with load cells (visible behind the top left and top right clamps). The blade is clamped to the load cell (not shown) of a uniaxial testing machine and is perpendicular to the plane of the skin specimen. Forces and displacements recorded by the load cell and actuator to which the blade is attached are analysed to characterise stab penetrations. The simulant material for underlying tissue is removed for clarity.*

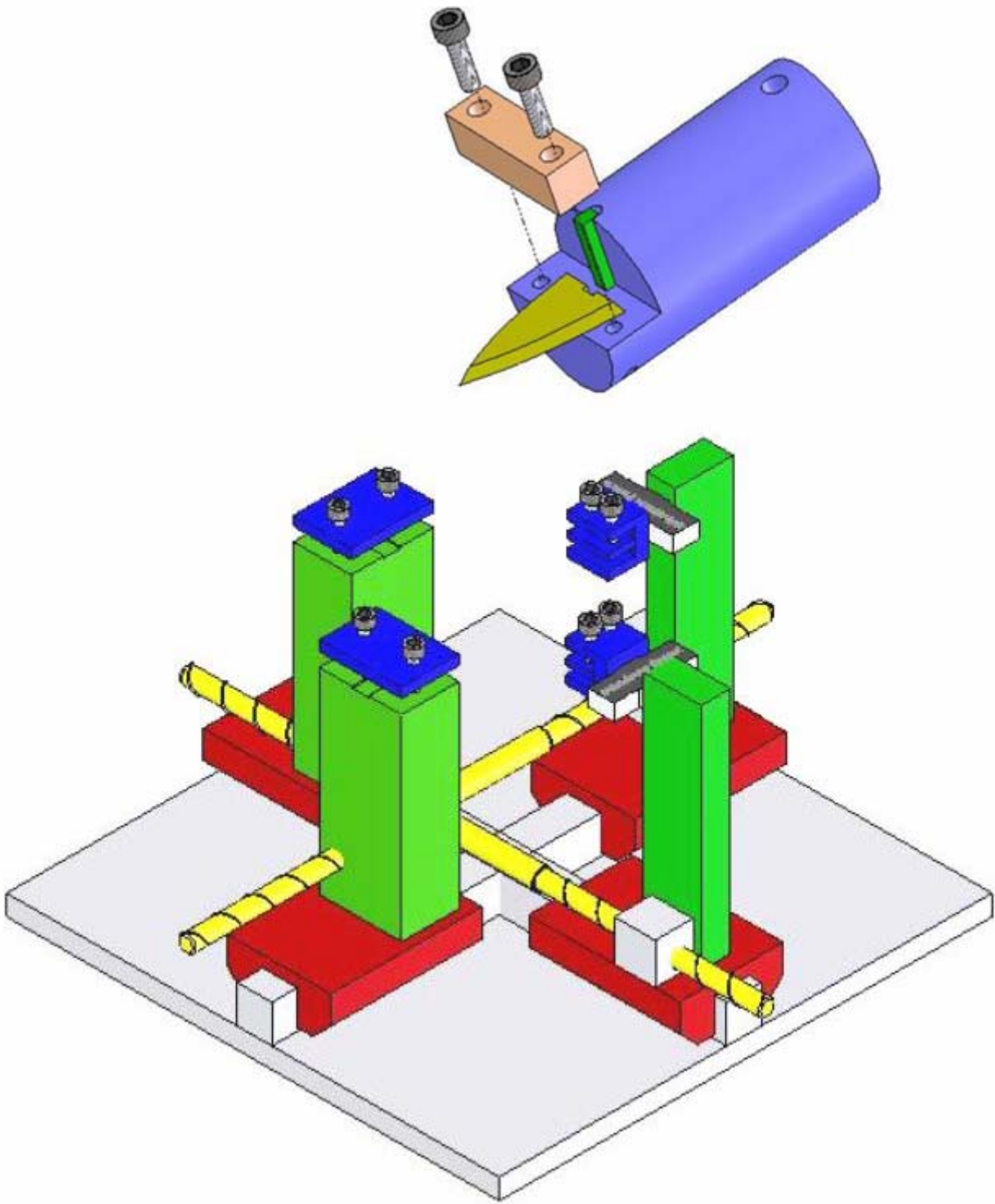

*Figure 4: Schematic of test arrangement showing (top) clamping of blade and (bottom) adjustable stage for clamping cruciform surrogate skin sample under biaxial tension with two pairs of clamps with load cells (seen immediately behind top right and bottom right clamps).*

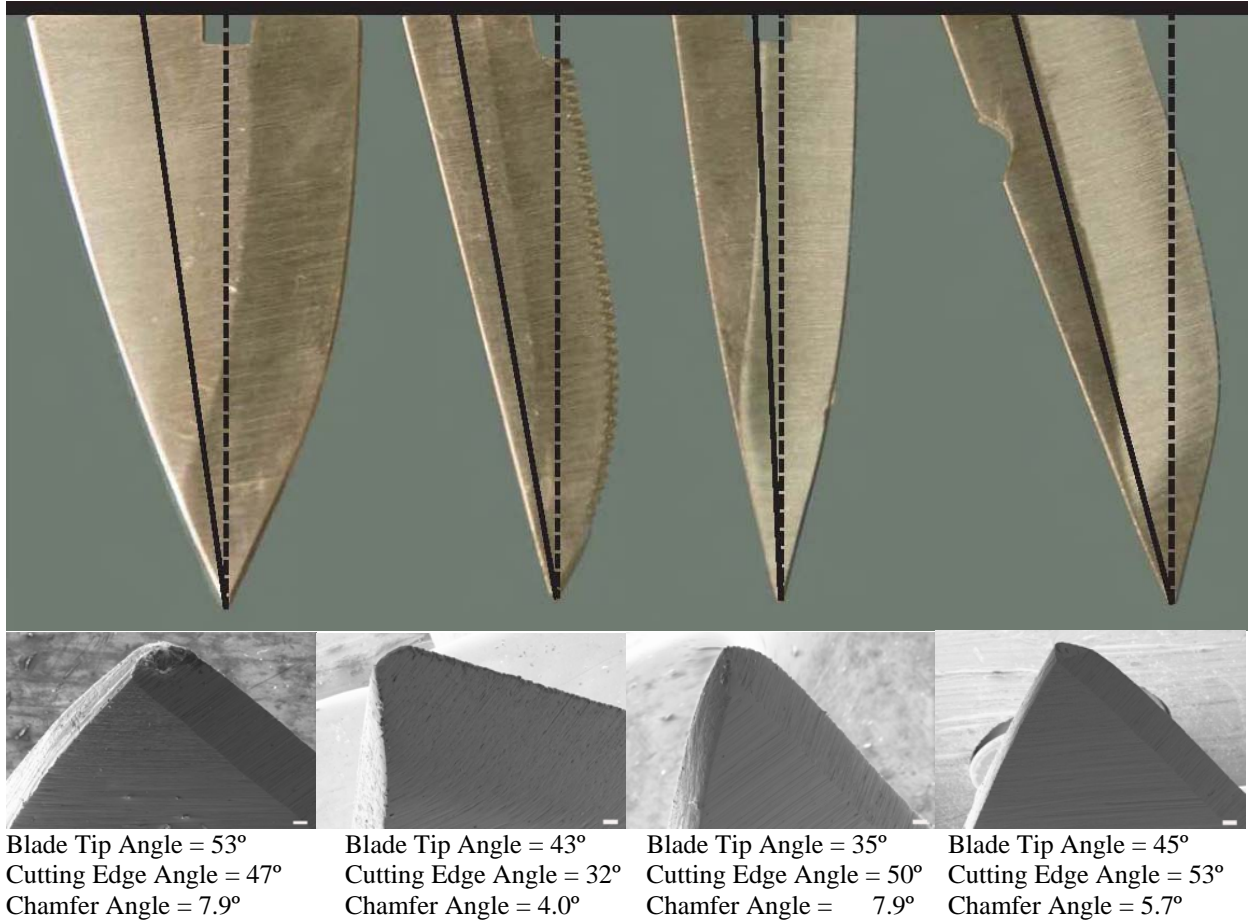

Blade Tip Angle = 53º  
Cutting Edge Angle = 47º  
Chamfer Angle = 7.9º

Blade Tip Angle = 43º  
Cutting Edge Angle = 32º  
Chamfer Angle = 4.0º

Blade Tip Angle = 35º  
Cutting Edge Angle = 50º  
Chamfer Angle =     7.9º

Blade Tip Angle = 45º  
Cutting Edge Angle = 53º  
Chamfer Angle = 5.7º

*Figure 5: Sectioned blades with locator notches from examples of each of the four knives examined, namely, cook's knife, kitchen knife, carving knife and utility knife (left to right; top photo) and corresponding scanning electron micrograph (SEM; bar markers correspond to 100μm) images of blade tip and point regions (bottom). The knives in the top photograph are perpendicular to plane of skin specimens (i.e., 0º), as indicated by dashed bisector line. The corresponding solid lines define axes from the point to the butt of the knife. The angle between the solid line and dashed line is 8º, 10º, 2.5º and 15º for the cook's, kitchen, carving and utility knives, respectively. The notch in the spine of the utility knife was machined to locate the blade against the screw in the blade clamp (c.f. Fig. 4).*

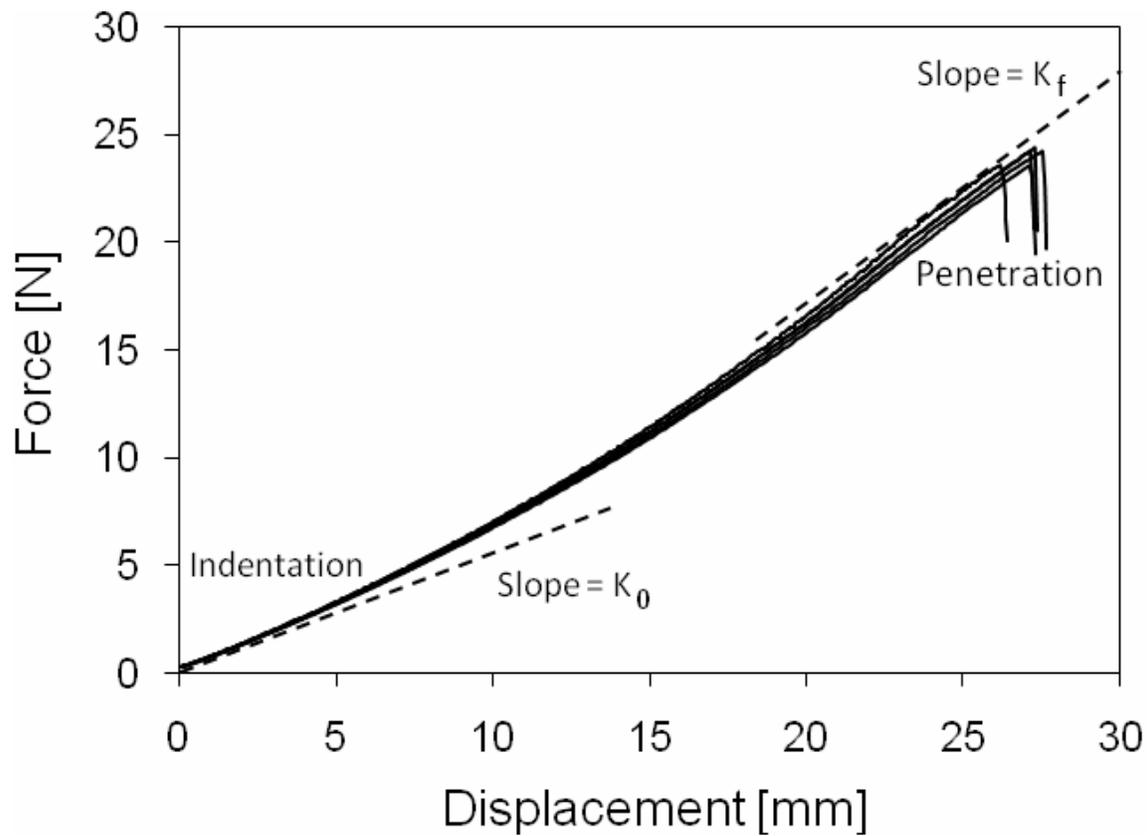

*Figure 6: Cook's knife stab test results from five experiments under 'reference' conditions, i.e., 4mm thick skin with no underlying tissue and equi-biaxial tension of 10N in both arms of the cruciform polyurethane skin simulants. Stab speed was 50mm/min and blade was aligned perpendicular to skin (i.e., angle = 0º). The dashed lines illustrate how initial and final tangent moduli for stiffnesses $K_0$ and $K_f$ respectively are calculated.*

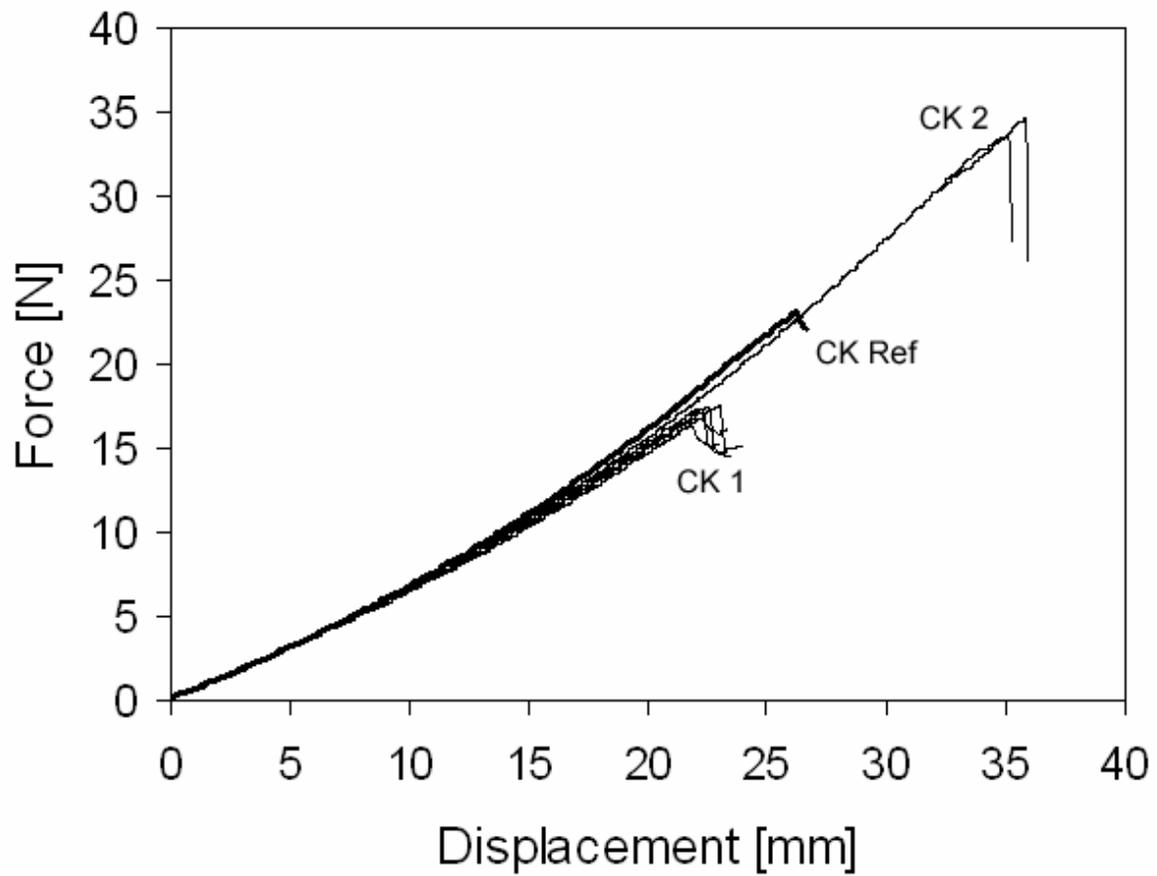

*Figure 7: Cook's knife stab test results under reference conditions: 4mm thick skin specimens without underlying tissue under equi-biaxial tension of 10N at 50mm/min at 0º. The average of the reference data of Fig. 6 is shown by the single heavy line (labelled 'CK Ref'). Data from two sets of at least three tests using other new and nominally identical cook's knives are indicated by 'CK 1' and 'CK 2'.*

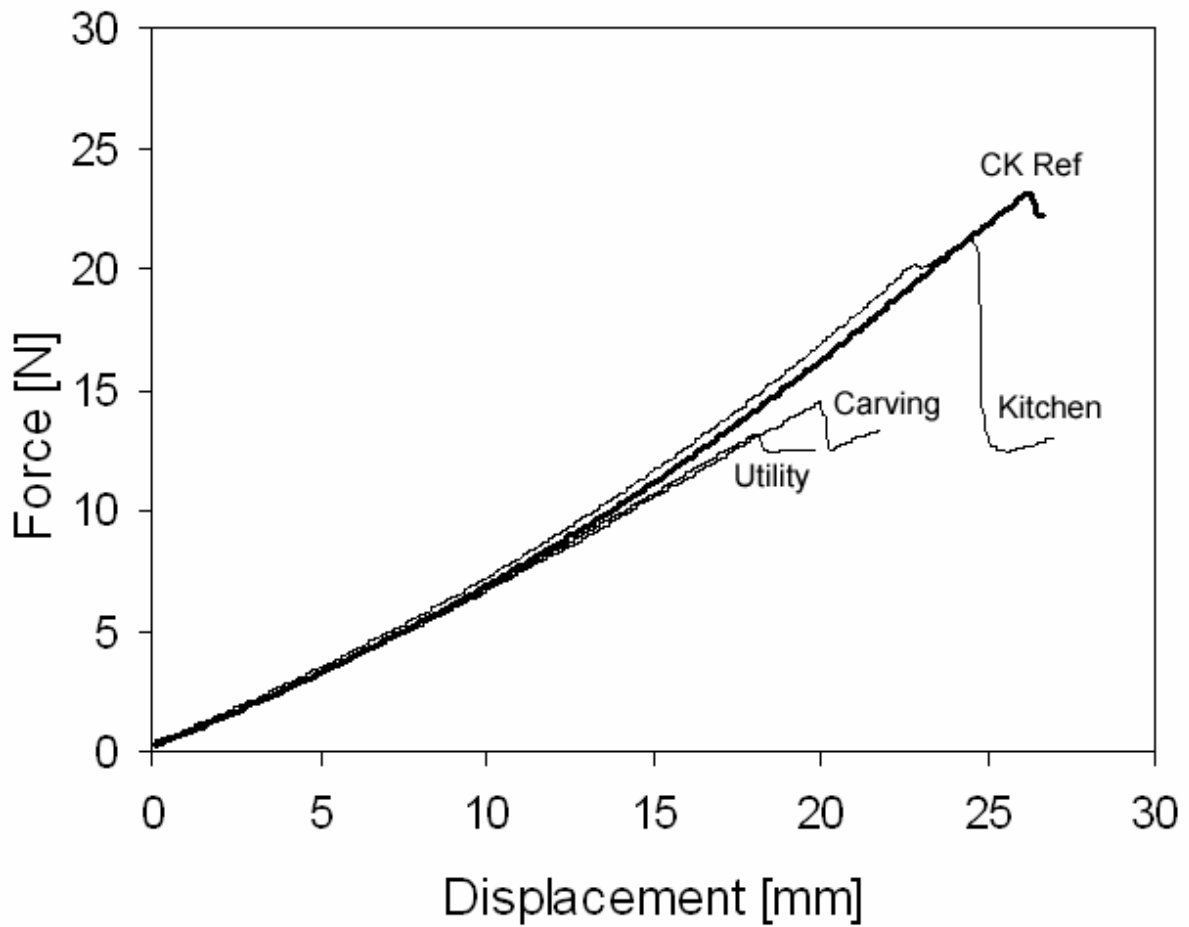

*Figure 8: Stab test results for all knives under reference conditions, i.e., 4mm thick skin specimens, 0º angles of knife orientation, no underlying tissue, equi-biaxial tension of 10N; knife speed of 50mm/min. The cook's knife (average of reference data in Fig. 6) is shown by heavy line whereas the utility, carving and kitchen knife curves are the average of at least three test repetitions each.*

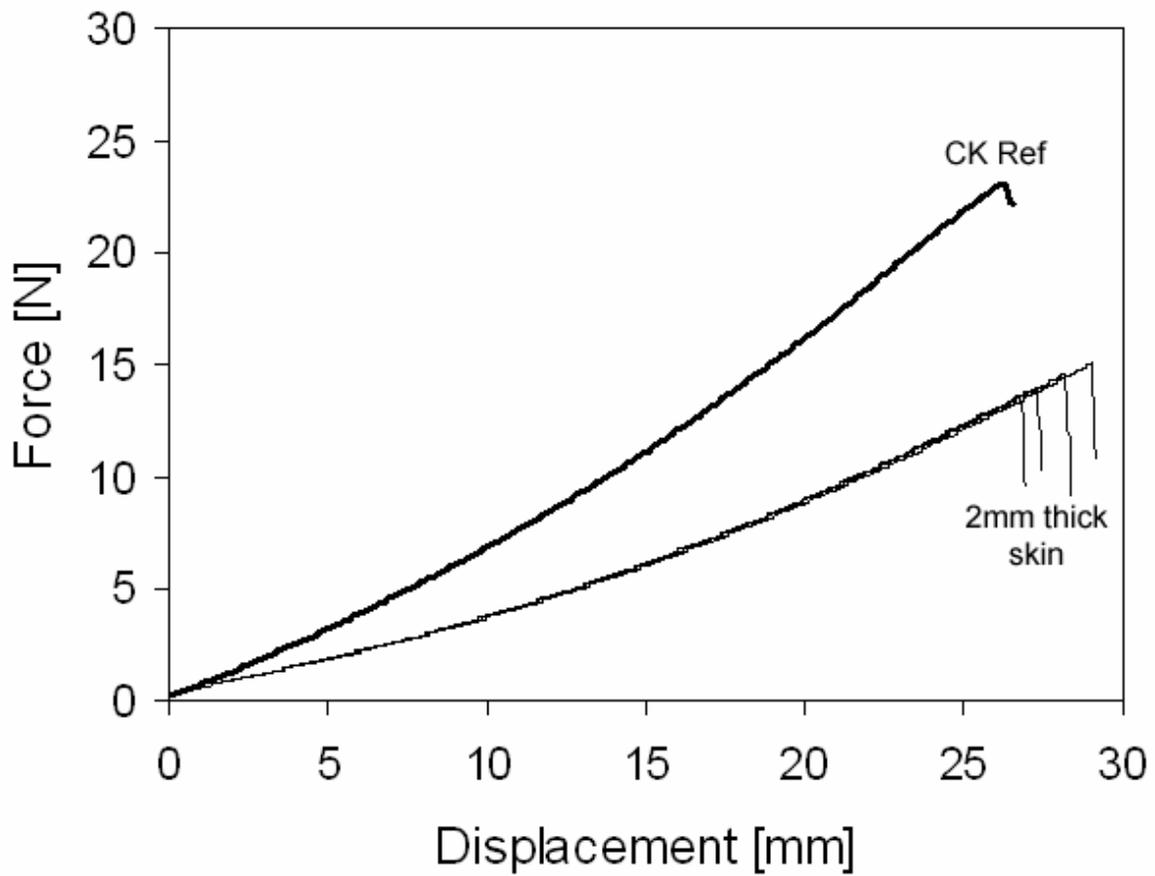

*Figure 9: Stab test results for the same cook's knife penetrating 4mm (average of reference data in Fig. 6 shown above by single heavy line) and 2mm (light lines represent four different tests) thick skin specimens. No underlying tissue was used; equi-biaxial tension of 10N was applied; knife speed and orientation were 50mm/min and 0º, respectively.*

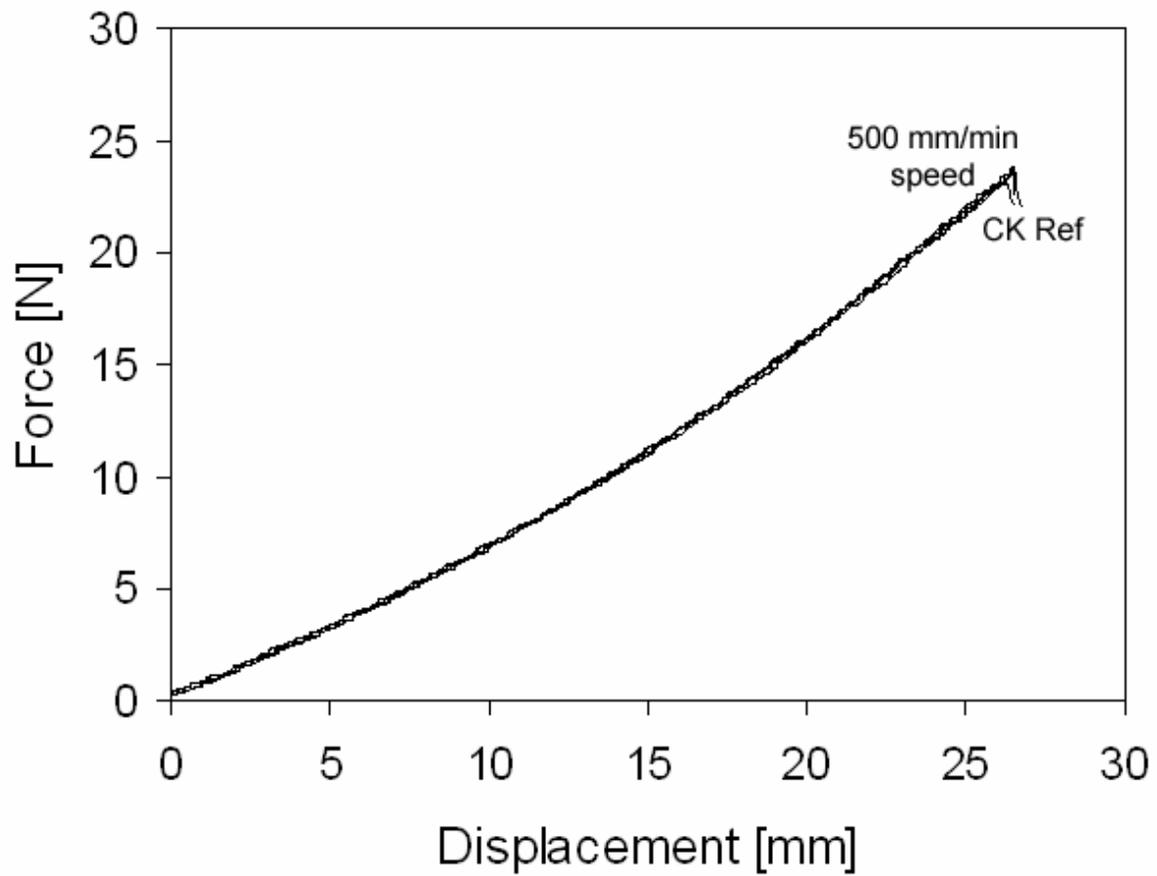

*Figure 10: Stab test results for the same cook's knife penetrating 4mm thick skin specimens at knife speeds of 50 mm/min (average of reference data in Fig. 6 shown above by single heavy line) and 500 mm/min (the three separate light lines represent three different tests). No underlying tissue was used; equi-biaxial tension of 10N was applied; knife orientation was 0º.*

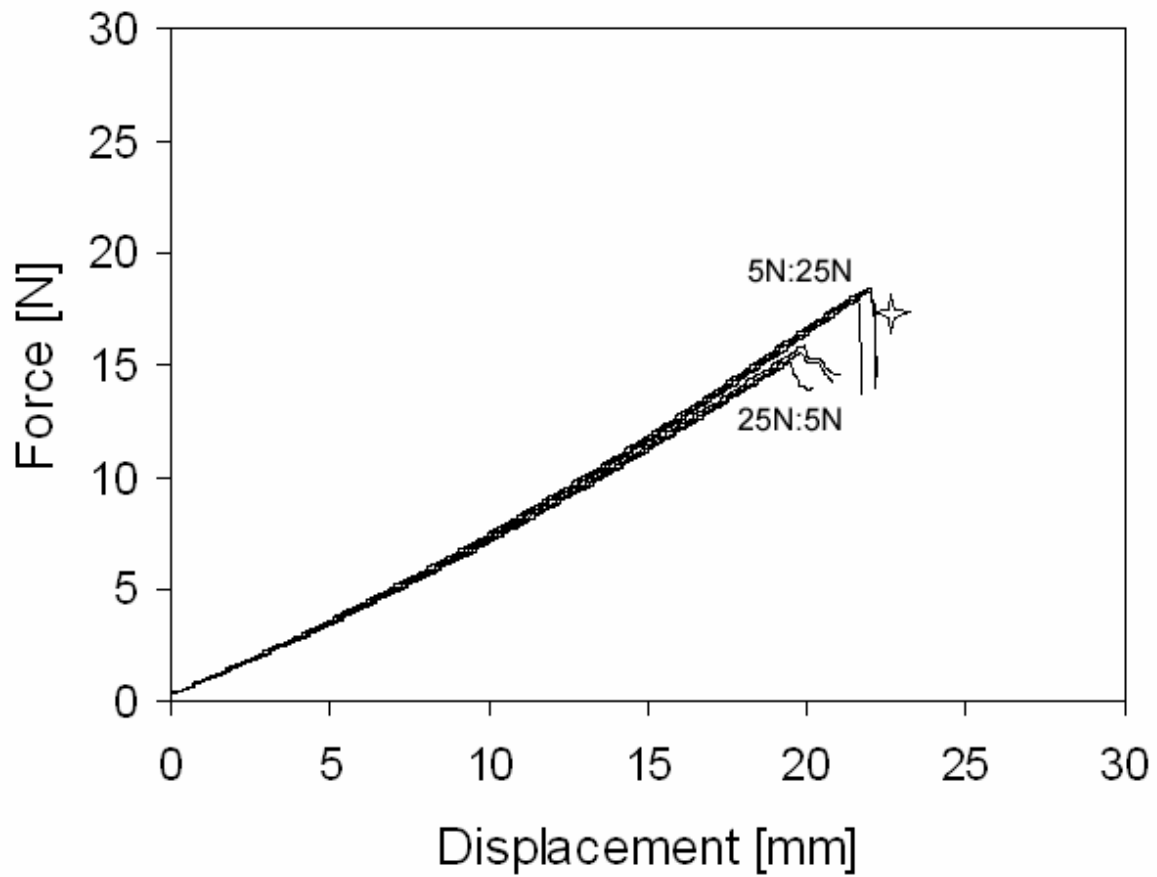

*Figure 11: Stab test results for the same cook's knife penetrating 4mm thick skin specimens under different and unequal levels of in-plane biaxial tension; knife speed and orientation were 50mm/min and 0º, respectively. Three tests are under biaxial tension of 5:25N (5N parallel to plane of blade and 25N perpendicular to blade) while three separate tests are under biaxial tension of 25:5N (25N parallel to plane of blade and 5N perpendicular to blade). Note that a new and nominally identical cook's knife ('CK 1' of Fig 7) was used for all of these tests. The star identifies the average $P_{max}$ and $\delta_{max}$ of Fig. 7, which correspond to tension of 10N:10N.*

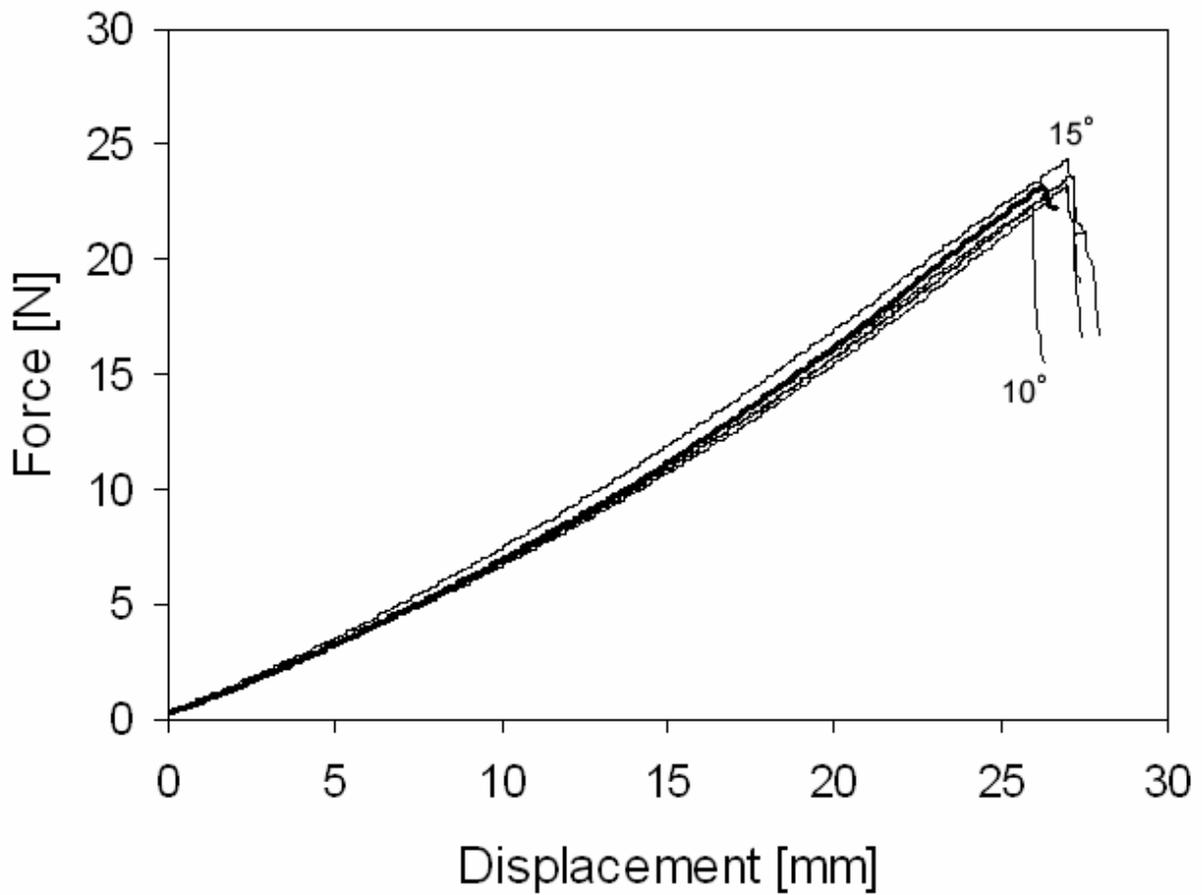

*Figure 12a: Stab test results for the same cook's knife penetrating 4mm thick skin specimens under different positive angles (1, 2, 3, 4, 5, 10 and 15°) of knife orientation (measured towards spine). No underlying tissue was used; equi-biaxial tension of 10N was applied; knife speed was 50mm/min. The average of reference data in Fig. 6 is shown in both cases by single heavy line.*

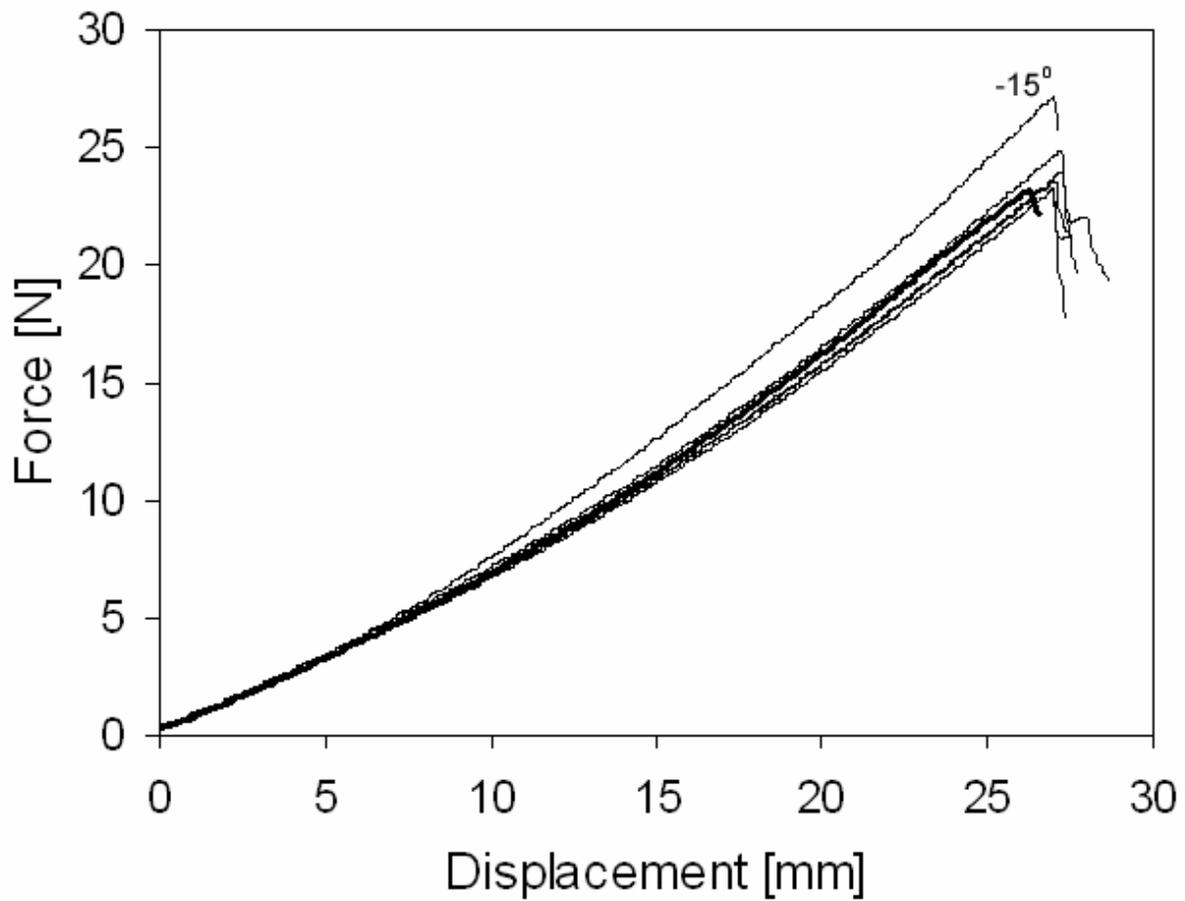

*Figure 12b: Stab test results for the same cook's knife penetrating 4mm thick skin specimens under different negative angles (-1, -2, -3, -4, -5, -10 and -15º) of knife orientation (measured towards edge). No underlying tissue was used; equi-biaxial tension of 10N was applied; knife speed was 50mm/min. The average of reference data in Fig. 6 is shown in both cases by single heavy line.*

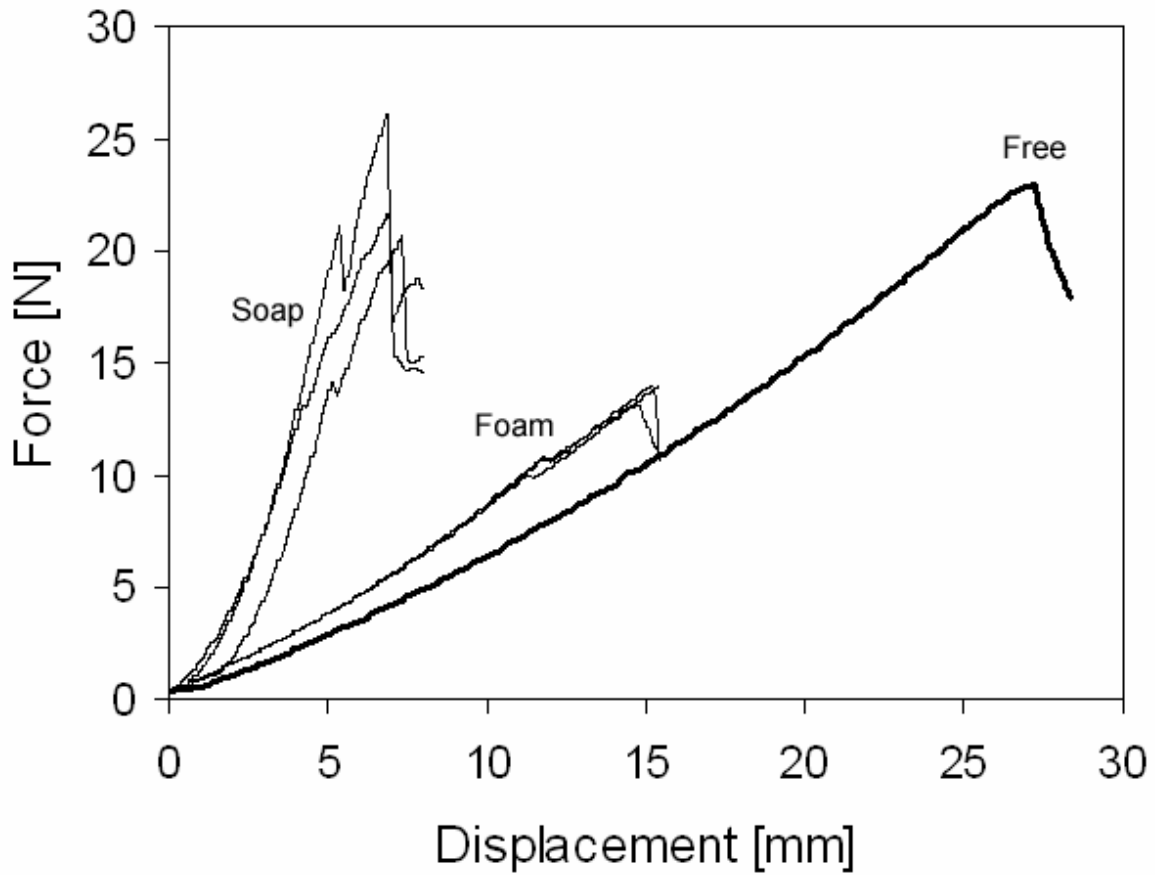

*Figure 13a: Stab test results for the same kitchen knife penetrating 4mm thick skin specimens on top of different underlying tissue. Ballistic soap provides stiffer results than compliant foam, while the least stiff results occur in the absence of any underlying tissue. The second peaks in the "soap" data correspond to the serrated section of the kitchen knife entering the tissue simulants. Equi-biaxial tension of 10N was applied; knife speed and orientation were 50mm/min and 0º, respectively.*

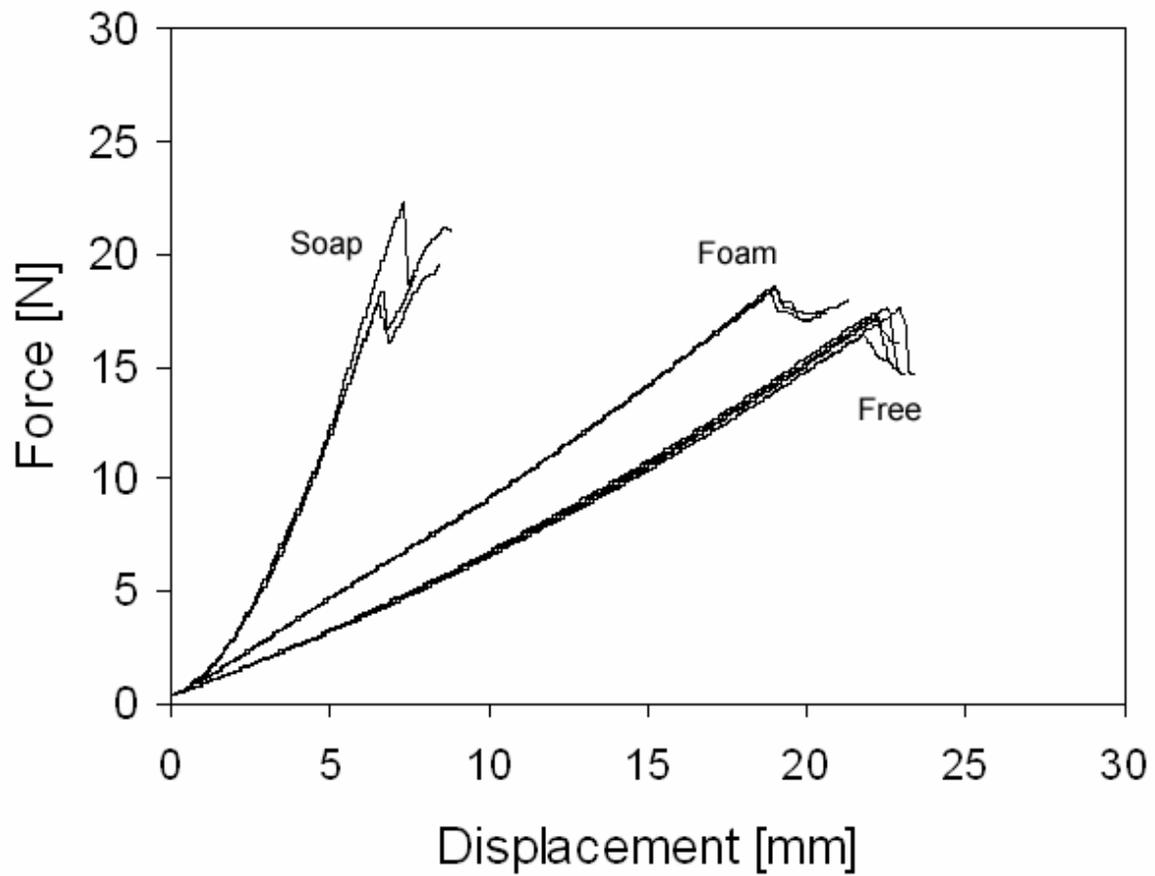

*Figure 13b: Stab test results for the same cook's knife('CK 1' of Fig. 7 penetrating 4mm thick skin specimens on top of different underlying tissue. Ballistic soap provides stiffer results than compliant foam, while the least stiff results occur in the absence of any underlying tissue. Equi-biaxial tension of 10N was applied; knife speed and orientation were 50mm/min and 0º, respectively.*

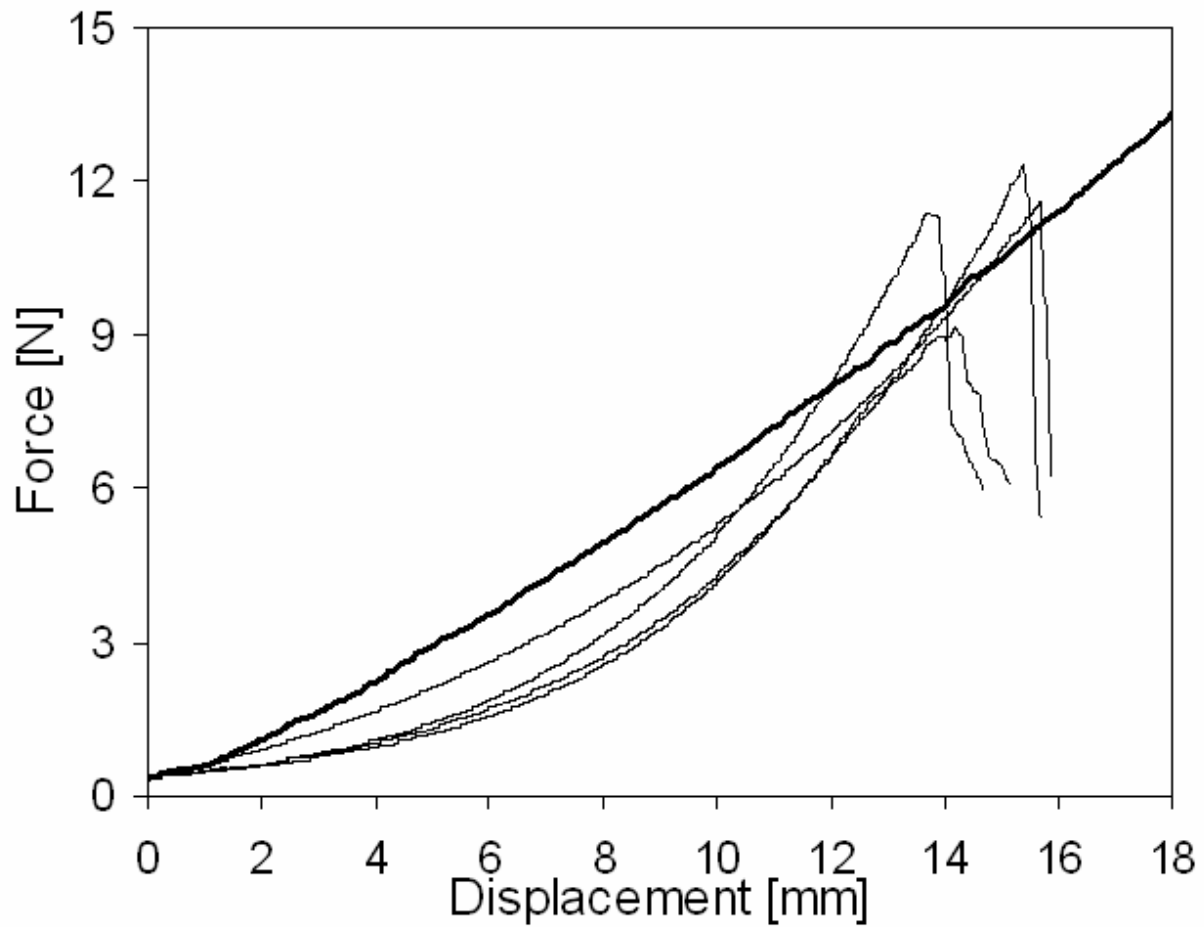

*Figure 14a: Stab test results for kitchen knife penetrating prepared specimens of leather. Equi-biaxial tension of 10N was applied; knife speed and orientation were 50mm/min and 0º, respectively, and no underlying tissue was present in either case. The heavy lines correspond to knives penetrating 4mm thick PU specimens under the same conditions. Note that the calf's leather used for each set of tests (Fig. 14a Vs 14b) had been cured differently and thus had slightly different mechanical properties.*

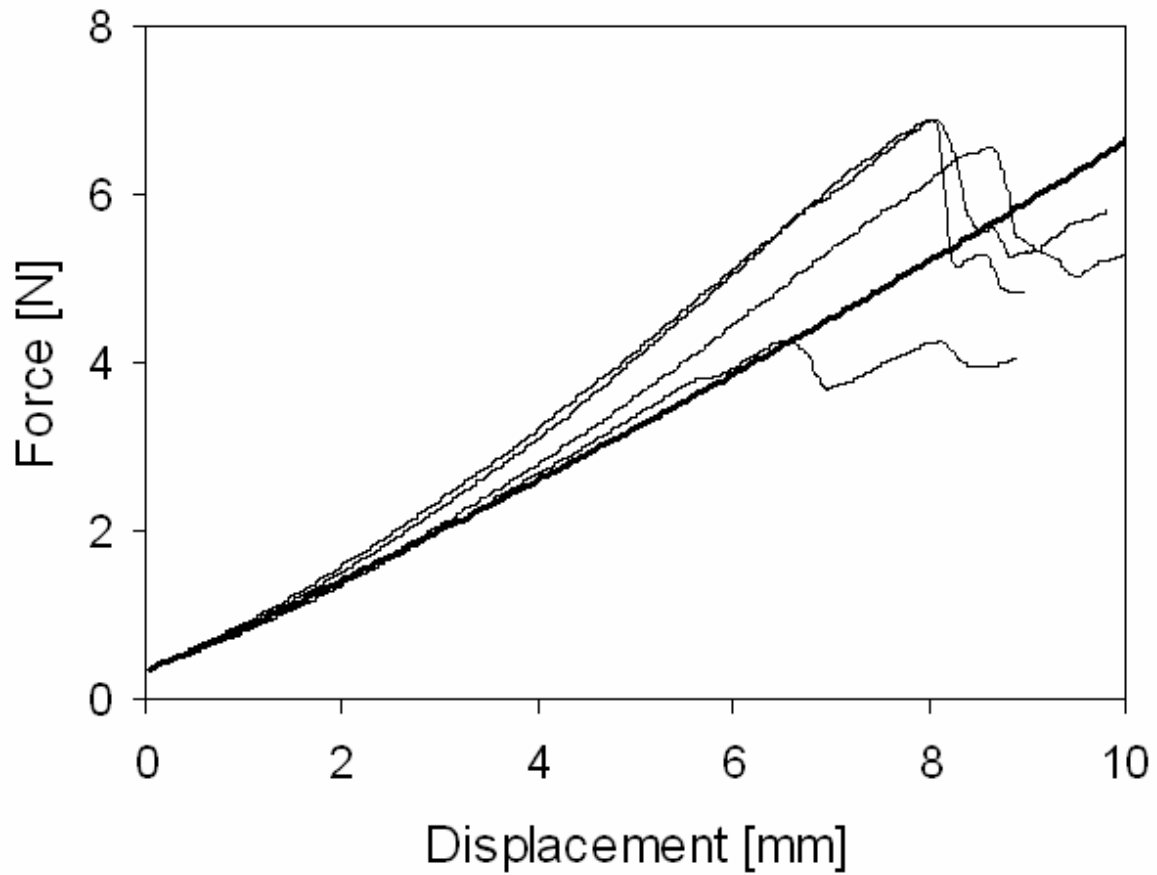

*Figure 14b: Stab test results for cook's knife ('CK 1') penetrating prepared specimens of leather. Equi-biaxial tension of 10N was applied; knife speed and orientation were 50mm/min and 0º, respectively, and no underlying tissue was present in either case. The heavy lines correspond to knives penetrating 4mm thick PU specimens under the same conditions. Note that the calf's leather used for each set of tests (Fig. 14a Vs 14b) had been cured differently and thus had slightly different mechanical properties.*